\documentclass[smallextended]{svjour3} 
\usepackage[T1]{fontenc}
\usepackage[latin1]{inputenc}
\usepackage{graphicx}
\usepackage{amssymb}
\usepackage{amsmath}
\usepackage{longtable}
\usepackage{wasysym}
\usepackage{booktabs}
\usepackage{natbib}
\usepackage[usenames]{color}
\usepackage{pdflscape}
\usepackage{rotating}
\usepackage{float}
\usepackage{soul}
\usepackage{xcolor, colortbl}
\definecolor{Gray}{gray}{0.9}
\usepackage{etoolbox}
\usepackage[letterpaper,margin=2.5cm]{geometry}
\usepackage{background}
\usepackage{setspace}
\AtBeginEnvironment{tabular}{\singlespacing}
\AtBeginEnvironment{longtable}{\singlespacing}

\backgroundsetup{contents={}}

\usepackage[pdftex,colorlinks,linkcolor=blue,anchorcolor=blue,urlcolor=blue,citecolor=blue]{hyperref}

\newtoggle{supplement}\toggletrue{supplement} 
\newtoggle{removemanunobox} \toggletrue{removemanunobox}

\newcommand{\startsupplement}{
\clearpage \newpage
\setcounter{page}{1} \renewcommand{\thepage}{S-\arabic{page}}
\setcounter{figure}{0} \renewcommand{\thefigure}{S-\arabic{figure}}
\setcounter{equation}{0} \renewcommand{\theequation}{S-\arabic{equation}}
\setcounter{table}{0} \renewcommand{\thetable}{S-\arabic{table}}
\setcounter{section}{0} \renewcommand{\thesection}{S-\arabic{section}}}

\iftoggle{removemanunobox}
{
\makeatletter
\renewcommand\makeheadbox{{%
\hbox to0pt{\vbox{\baselineskip=10dd\hbox
to\hsize{\kern3pt\vbox{\kern12pt
\hbox{\ }
\hbox{}
}}}%
}}}
\makeatother
}

\journalname{Climatic Change}

\begin{document}


\title{Coastal flood implications of 1.5 $^\circ$C, 2.0 $^\circ$C, and 2.5 $^\circ$C temperature stabilization targets in the 21st and 22nd century}
\author{D.J. Rasmussen, Klaus Bittermann, Maya K. Buchanan, Scott Kulp, Benjamin H. Strauss, Robert E. Kopp, Michael Oppenheimer}

\institute{D.J. Rasmussen, Maya K. Buchanan and Michael Oppenheimer \at Woodrow Wilson School of Public \& International Affairs, Princeton University, Princeton, NJ, USA.  \email{dj.rasmussen@princeton.edu}
\and Klaus Bittermann \at Department of Earth and Ocean Sciences, Tufts University, Medford, MA, USA and Potsdam Institute for Climate Impact Research, Potsdam, Germany
\and Scott Kulp and Benjamin H. Strauss \at Climate Central, Princeton, NJ, USA
\and Michael Oppenheimer \at Department of Geosciences, Princeton University, Princeton, NJ, USA
\and Robert E. Kopp \at Department of Earth \& Planetary Sciences and Institute of Earth, Ocean, \& Atmospheric Sciences, Rutgers University, New Brunswick, NJ, USA.}





\date{Draft: \today}
\maketitle 

\begin{abstract}

Sea-level rise (SLR) is magnifying the frequency and severity of coastal flooding. The rate and amount of global mean sea-level (GMSL) rise is a function of the trajectory of global mean surface temperature (GMST). Therefore, temperature stabilization targets (e.g., 1.5 $^\circ$C and 2.0 $^\circ$C of warming above pre-industrial levels, as from the Paris Agreement) have important implications for coastal flood risk. Here, we assess differences in the return periods of coastal floods at a global network of tide gauges between scenarios that stabilize GMST warming at 1.5 $^\circ$C, 2.0 $^\circ$C, and 2.5 $^\circ$C above pre-industrial levels. We employ probabilistic, localized SLR projections and long-term hourly tide gauge records to construct estimates of the return levels of current and future flood heights for the 21st and 22nd centuries. By 2100, under 1.5 $^\circ$C, 2.0 $^\circ$C, and 2.5 $^\circ$C GMST stabilization, median GMSL is projected to rise 47 cm with a \emph{very likely} range of 28--82 cm (90\% probability), 55 cm (\emph{very likely} 30--94 cm), and 58 cm (\emph{very likely} 36--93 cm), respectively. As an independent comparison, a semi-empirical sea level model calibrated to temperature and GMSL over the past two millennia estimates median GMSL will rise within < 13\% of these projections. By 2150, relative to the 2.0 $^\circ$C scenario, GMST stabilization of 1.5 $^\circ$C inundates roughly 5 million fewer inhabitants that currently occupy lands, including 40,000 fewer individuals currently residing in Small Island Developing States. Projected changes to the frequency of current 10-, 100-, and 500-year flood levels are quantified using flood amplification factors that incorporate uncertainty in both historical flood return periods and local SLR. Relative to a 2.0 $^\circ$C scenario, the reduction in the amplification of the frequency of the 100-yr flood arising from a 1.5 $^\circ$C GMST stabilization is greatest in the eastern United States and in Europe, with flood frequency amplification being reduced by about half.

\end{abstract}

\section{Introduction}




Coastal flooding is a hazard that threatens both life and property. The height of a coastal flood is determined by the combined height of the astronomical tide and storm surge (i.e., the storm tide) and the mean sea level at the time of the event. Rising mean sea levels are already magnifying the frequency and severity of coastal floods \citep{Buchanan2017,Sweet2014} and by the end of the century, coastal floods figure to be among the costliest impacts of climate change in some regions \citep{Hsiang2017,Diaz2016}. Sea-level rise (SLR) is expected to permanently inundate low-lying geographic areas \citep{Marzeion2014,Strauss2015a}, but these locations will first experience decreases in the return periods of flood events \citep[e.g.,][]{Hunter2012,Sweet2014}. 

The rate of global mean sea-level (GMSL) rise depends on the trajectory of global mean surface temperature (GMST; \citealp{Rahmstorf2007,Kopp2016,Vermeer2009}), with the long-term committed amount of GMSL largely determined by the stabilized level of GMST \citep{Levermann2013}. Thus, the management of GMST has important implications for regulating future GMSL \citep{Schaeffer2012}, and consequently the frequency and severity of coastal floods. However, GMST stabilization does not imply stabilization of all climate variables. Under stabilized GMST, GSML is expected to continue to rise for centuries, due to the long residence time of anthropogenic CO$_{2}$, the thermal inertia of the ocean, and the slow response of large ice sheets to forcing \citep{Clark2016,Levermann2013,Held2010}. For instance, \citet{Schaeffer2012} found that a 2.0 $^\circ$C GMST stabilization would lead to a GMSL rise (relative to 2000) of 0.8 m by 2100 and > 2.5 m by 2300, but if the GMST increase were held below 1.5 $^\circ$C, GMSL rise at the end of the 23rd century would be limited to $\sim$1.5 m. These findings suggest that selection of climate policy goals could have critical long-term consequences for the impacts of future SLR and coastal floods \citep{Clark2016}. 

The Paris Agreement seeks to stabilize GMST by limiting warming to ``well below 2.0 $^\circ$C above pre-industrial levels'' and to further pursue efforts to ``limit the temperature increase to 1.5 $^\circ$C above pre-industrial levels'' \citep{UNFCCC2015a}. However, a recent literature review under the United Nations Framework Convention on Climate Change (UNFCCC) found the notion that ``up to 2.0 $^\circ$C of warming is considered safe, is inadequate'' and that ``limiting global warming to below 1.5 $^\circ$C would come with several advantages'' \citep{UNFCCC2015b}. The advantages and disadvantages of each GMST target as they relate to coastal flooding have not been quantified. This is critical as > 625 million people currently live in low-elevation coastal zones, and population growth is expected in these areas \citep{Neumann2015}. Examining the short- and long-term flood hazard implications of 1.5 $^\circ$C and 2.0 $^\circ$C GMST stabilization scenarios, as others have recently done for other climate impacts \citep[e.g.][]{Schleussner2016,Schleussner2016b,Mitchell2017,Mohammed2017}, may better inform the policy debate regarding the selection of GMST goals.

In this study, we employ probabilistic, localized SLR projections to assess differences in the frequency of extreme coastal floods across 1.5 $^\circ$C, 2.0 $^\circ$C, and 2.5 $^\circ$C GMST stabilization scenarios at a global network of 194 tide gauges. We use long-term hourly tide gauge records and extreme value theory to estimate present and future return periods of flood events. We extend our analysis through the 22nd century to account for continuing SLR in order to inform multi-century planning and infrastructure investments. Lastly, we assess differences in the exposure of current populations to future SLR under  1.5 $^\circ$C, 2.0 $^\circ$C, and 2.5 $^\circ$C GMST stabilizations.

Various approaches have been used to project GMSL under GMST targets. For instance, \citet{Jevrejeva2016} estimate future local SLR under a GMST increase of 2 $^\circ$C using an RCP8.5 GMST trajectory that passes through 2 $^\circ$C of warming by mid-century, but this approach likely underestimates SLR relative to a scenario that achieves 2 $^\circ$C GMST stabilization by 2100 as it neglects the time-lagged, integrated response of the ocean and cryosphere to warming \citep{Clark2016}. More generally, studies that condition future flood projections on the Representative Concentration Pathways (RCPs) may be insufficient for assessing the costs and benefits of climate policy scenarios, such as GMST stabilization targets \citep[e.g., \emph{Section 13.7.2.2} of][]{Church2013,Buchanan2017,Hunter2012,Tebaldi2012}. The RCPs are designed to be representative of a range of emissions scenarios that result in prescribed anthropogenic radiative forcings by 2100 relative to pre-industrial conditions (e.g., 8.5 W$m^{-2}$ for RCP8.5). They are not representative of a specific emissions trajectory, climate policy (e.g., GMST target), or socioeconomic and technological change \citep{Moss2010,vanVuuren2011}. 

Semi-empirical sea level (SESL) models \citep{Rahmstorf2012} can estimate future GMSL rise under various GMST scenarios \citep[e.g.,][]{Schaeffer2012,BittermanInRev}. Unlike their process-based counterparts \citep[e.g.,][]{Kopp2014}, SESL models do not explicitly model individual physical components of sea-level change. They are calibrated over a historical period using the observed statistical relationship between GMSL and a climate parameter (such as GMST). Assuming these relationships hold in the future, SESL models project the rate of GMSL change conditional upon a GMST pathway \citep[e.g.,][]{Rahmstorf2007,Vermeer2009,Kopp2016}. However, SESL models do not produce estimates of local SLR, which are necessary for local risk assessment and adaptation planning because local SLR can substantially differ from the global mean \citep{Milne2009}.

\section{Methods}

We project global and local sea level under 1.5 $^\circ$C, 2.0 $^\circ$C, and 2.5 $^\circ$C GMST stabilization targets using the component-based, probabilistic, local sea level projection framework from \citet[][henceforth K14]{Kopp2014}. We compare the resulting GMSL projections to those from the semi-empirical sea level (SESL) model of \citet{Kopp2016}. While SESL models cannot produce local projections of SLR, they can serve as a reference point for evaluating the consistency of process-based projections with historical temperature-GMSL relationships. The flow and sources of information used to construct the local SLR and GMSL projections using the K14 method is depicted in Fig. \ref{Sfig:flow_diag}A, while the flow of information used to generate the SESL projections is provided in Fig. \ref{Sfig:flow_diag}B. Local SLR projections from the K14 approach are combined with historical flood distributions to estimate future return periods of historical flood events (Fig. \ref{Sfig:flow_diag}A). 


\subsection{Component-based model approach: Global and local sea-level rise projections}

Global sea-level change does not occur uniformly. Dynamic ocean processes \citep{Levermann2005}, changes to temperature and salinity (i.e., steric processes), changes in the Earth's rotation and gravitational field associated with water-mass redistribution (e.g., land-ice melt; \citealp{Mitrovica2011}), and glacial isostatic adjustment (GIA;  \citealp{Farrell1976}) cause local sea levels to differ from the global mean. We model local sea level using the K14 framework, but make modifications to accommodate the stratification of Atmosphere-Ocean General Circulation Models (AOGCMs) and RCPs into groups that meet GMST stabilization targets (see Section \ref{sec:fccs}). AOGCM output from the Coupled Model Intercomparison Project (CMIP) Phase 5 archive \citep{Taylor2012} forced with the RCPs (to 2100) and their extensions (to 2300) are used for the following SLR components: global mean thermal expansion (TE), local ocean dynamics, and glacial ice contributions (GIC; \citealp{Marzeion2012}). Antarctic Ice Sheet (AIS) and the Greenland Ice Sheet (GIS) contributions are estimated using a combination of the Intergovernmental Panel on Climate Change's (IPCC) Assessment Report 5 (AR5) projections of ice sheet dynamics and surface mass balance (SMB) \citep[Table 13.5 in][]{Church2013} and expert elicitation of total ice sheet mass loss from \citet{Bamber2013}. As in AR5, ice sheet SMB contributions are represented as being dependent on the forcing scenario, while ice sheet dynamics are not. Here, we use AR5's AIS and GIS SMB contributions from RCP2.6 for both the 1.5 $^\circ$C and 2.0 $^\circ$C GMST scenarios and the RCP4.5 projection for the 2.5 $^\circ$C GMST scenario \citep[Table 13.5 in][]{Church2013}. A spatiotemporal Gaussian process regression model is used to estimate the long-term contribution from non-climatic factors such as tectonics and GIA. To generate probability distributions of global mean and local sea level at a global network of tide gauge sites (Table \ref{Stab:tgauges}) for each GMST scenario, we use 10,000 Latin hypercube samples of probability distributions of individual sea level component contributions.

\subsubsection{Approximating Global Temperature Stabilization with RCPs} \label{sec:fccs}

The RCP-driven experiments in the CMIP5 archive are not designed to inform the assessment of climate impacts from incremental temperature changes. As such, we construct alternative ensembles for 1.5 $^\circ$C, 2.0 $^\circ$C, and 2.5 $^\circ$C scenarios using CMIP5 output according to each AOGCM's 2100 GMST. Specifically, we create ensembles for 1.5 $^\circ$C, 2.0 $^\circ$C, and 2.5 $^\circ$C scenarios with AOGCMs that have a 2100 GMST increase (19-yr running average) of 1.5 $^\circ$C, 2.0 $^\circ$C, and 2.5 $^\circ$C ($\pm$ 0.25 $^\circ$C) relative to 1875--1900. Selection of the AOGCMs for each scenario ensemble are made irrespective of the AOGCM's RCP forcing. For model outputs that end in 2100, we extrapolate 19-yr running average GMST to 2100 based on the 2070--2090 trend. While we chose 2100 as the determining year for which AOGCMs are selected for each ensemble, it should be noted that Article 2 of the UNFCCC \citep{UNFCCC1992} does not require that GMST stabilization be achieved within a particular time frame. The Paris Agreement likewise does not specify a timeframe for GMST stabilization, though its goal of brining net greenhouse gas (GHG) emissions to zero in the second half of the 21st century implies a similar time frame for stabilization. We make the assumption that AOGCM outputs that end at 2100 either stay within the range of the target $\pm$ 0.25 $^\circ$C or fall below by any amount (i.e., undershoot). For AOGCMs that have GSMT output available after 2100, only those that undershoot the target are retained. However, we make an exception to this rule for the 2.5 $^\circ$C scenario ensemble in order to include AOGCMs for generating post-2100 projections. For RCP4.5 and RCP6, GMST stabilization should not occur before 2150, when greenhouse gas concentrations stabilize \citep{Meinshausen2011a} and so SLR projections after 2100 may not be representative of conditions under true GMST stabilization. 

The GMST trajectories and GMSL contributions from TE and glacial ice from selected CMIP5 models that are binned into 1.5 $^\circ$C, 2.0 $^\circ$C, and 2.5 $^\circ$C GMST categories are shown in Figs. \ref{fig:gmst} and \ref{Sfig:zostoga_gic}, respectively. For consistency with the K14 framework, which models 19-yr running averages of SLR relative to 2000, GMST is anomalized to 1991--2009 and then shifted upward by 0.72 $^\circ$C to account for warming since 1875--1900 \citep{Hansen2010,GISS2017}. Table \ref{Stab:model_inventory} lists the AOGCMs employed in each GMST scenario ensemble and the sea-level components used. Given the paucity of CMIP5 output after 2100, the range of TE and GIC contributions to SLR in the 22nd century is likely underestimated relative to the 21st century. 

\subsection{Global mean sea-level rise projections from a semi-empirical sea level model}

We generate estimates for GMSL for 2000--2100 using the SESL model from \citet{Kopp2016} driven with both GMST trajectories from CMIP5 models (Fig. \ref{fig:gmst}) and GMST trajectories from the reduced-complexity climate model MAGICC6 \citep[][as employed in \citealp{Rasmussen2016}]{Meinshausen2011b} for 2100 GMST targets of 1.5 $^\circ$C, 2.0 $^\circ$C, and 2.5 $^\circ$C  ($\pm$ 0.25 $^\circ$C) (Fig. \ref{Sfig:sesl_gmst}). The MAGICC6 GSMT trajectories are selected from all RCP-grouped projections using the same criteria as in Section \ref{sec:fccs}. The SESL model is calibrated to the Common Era temperature reconstruction from \citet{Mann2009} and the sea level reconstruction of \citet{Kopp2016}. The historical statistical relationship between temperature and the rate of sea-level change is assumed to be constant; not included are nonlinear physical processes or critical threshold events that could substantially contribute to SLR, such as ice sheet collapse \citep{Kopp2016b, Levermann2013}. Threshold behavior is partially incorporated in the K14 framework through expert assessments of future ice sheet melt contributions \citep{Bamber2013}, which may be one reason why the K14 framework produces higher estimates in the upper tail of the SLR probability distribution for 2100. 

\subsection{Flood frequency estimation}
\subsubsection{Historic flood return levels} \label{returncurves}

Extreme value theory is used with tide gauge observations to estimate the return levels of flood events of a given height, including those that occur less often, on average, than the length of the observational record \citep{Coles2001a,Coles2001b}. Following \citet{Tebaldi2012} and \citet{Buchanan2016,Buchanan2017}, we employ a generalized Pareto distribution (GPD) and a peaks-over-threshold approach to estimate the return periods of historical flood events at tide gauges. The GPD describes the probability of a given flood height conditional on an exceedance of the GPD threshold. Here, we use as the threshold the 99th percentile of daily maximum sea levels, which is generally both above the highest seasonal tide and balances the bias-variance trade-off in the GPD parameter estimation\footnote{If too low of a GPD threshold is chosen, more observations than those exclusively in the tail of the GPD distribution might end up being included in the parameter calculation, causing bias. If too high of a GPD threshold is chosen, then too few observations may be incorporated in the estimation of distribution parameters leading to greater variance, relative to a case that uses more observations.} \citep{Tebaldi2012}. The number of annual exceedances of the GPD threshold is assumed to be Poisson distributed with mean $\lambda$. The GPD parameters are estimated using the method of maximum likelihood with tide gauge observations referenced to Mean Higher High Water (MHHW)\footnote{Here defined as the average level of high tide over the last 19-years in each tide gauge record, which is different from the current U.S. National Tidal Datum Epoch of 1983--2001.} above the 99th percentile in each tide gauge's record (\emph{see Supporting Information}). Uncertainty in the GPD parameters is calculated from their covariance and is sampled using Latin hypercube sampling of a 1000 normally distributed GPD parameter pairs. For a given tide gauge, the annual expected number of exceedances of flood level $z$ is given by $N$($z$):

\begin{equation}
N(z)= \begin{cases}
\lambda\left ( 1+\frac{\xi(z-\mu)}{\sigma } \right )^{-\frac{1}{\xi }} &\mbox{for } \xi \neq 0\\ 
\lambda\text{ exp}(-\frac{z-\mu}{\sigma})&\mbox{for } \xi =0\\
\end{cases}
\end{equation}

\noindent where the shape parameter ($\xi$) governs the curvature and upward statistical limit of the flood return curve, the scale parameter ($\sigma$) characterizes the variability in the exceedances caused by the combination of tides and storm surges, and the location parameter ($\mu$) is the threshold water-level above which return-levels are estimated with the GPD. Meteorological and hydrodynamic differences between sites gives rise to differences in the shape parameter ($\xi$). Flood frequency distributions with $\xi$ > 0 are ``heavy tailed'', due to a high frequency of extreme flood events (e.g., tropical and extra-tropical cyclones). Distributions with $\xi$ < 0 are ``thin tailed'' and have a statistical upper bound on extreme flood levels. Events that occur between $\lambda$ and 182.6/year (i.e., exceeding MHHW half of the days per year) are modeled with a Gumbel distribution as they are outside of the domain of the GPD.

\subsubsection{Flood frequency amplification factors}

The amplification factor (AF) quantifies the increase in the expected frequency of historical flood events (e.g., the 100-yr flood) due to SLR \citep{Buchanan2017,Hunter2012,Church2013}. Due to variation in the local storm climate and hydrodynamics, the height of flood events are unique to each location (SI Fig. \ref{Sfig:histFloodMaps}). The calculation of the expected AF includes both the uncertainty in the estimates of the return periods of historical flood events and uncertainty in SLR projections. Following \citet{Buchanan2017}, we define the expected flood amplification factor $\text{AF}(z)$ for flood events with height $z$ as the ratio of the expected number of flood events after including uncertain SLR ($\delta$) to the historical expected number of flood events:

\begin{equation}
\text{AF}(z) = \frac{\text{E}[N(z-\delta)]}{N(z)}
 \end{equation}
 
\noindent The flood AFs reported in this study should not be directly interpreted as changes in flood frequency. What constitutes a future flood event critically depends on the future level of coastal protection and adaptation efforts, both of which are unknown for most locations.

\subsubsection{Assessment of population exposure}
Following the methods used in \citet{Kopp2017}, we assess the current population exposed to permanent inundation from GMSL under each GMST stabilization scenario. We caution that this is not a measure of future population exposure, which will depend upon both population growth and the dynamic response of the population to rising sea levels, but is instead intended to provide a summary metric of the human significance of different sea levels. We use a 1-arcsec SRTM 3.0 digital elevation model from NASA \citep{NASA2013} that is referenced to local MHHW levels and this study's local SLR projection grids. Projected inundation areas are intersected with current national population \citep{Bright2011} and national boundary data \citep{Hijmans2012}. For each GMST target, the population exposed is assessed at the 50th, 5th, and 95th percentile local SLR projection. Further details are provided in the Supplementary Information of \citet{Kopp2017}.

\section{Results} 

\subsection{Global mean sea-level rise}

\begin{figure}[h]
\centering
  \includegraphics[width=0.99\textwidth]{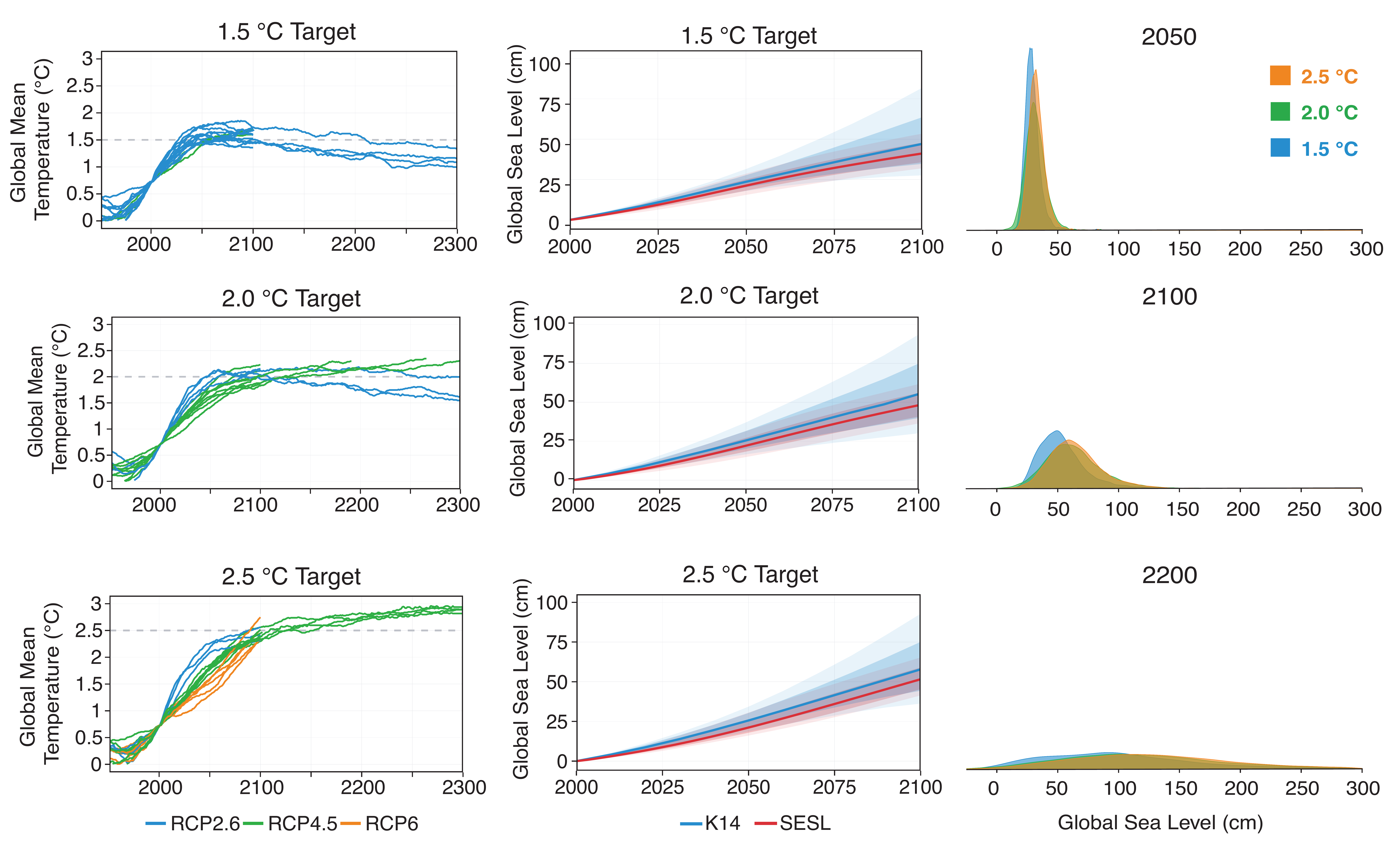}   
  \caption{\textbf{Left Column:} Global mean surface temperature (GMST) trajectories from CMIP5 models (1950--2300) that have a 19-year running average 2100 GMST of 1.5 $^\circ$C, 2.0 $^\circ$C, and 2.5 $^\circ$C $\pm$ 0.25$^\circ$C (relative to 1875--1900) (blue = RCP2.6, green = RCP4.5, orange = RCP6). GMST is anomalized to 1991--2009 and shifted up by 0.72$^\circ$C to account for warming since 1875--1900 \citep{Hansen2010,GISS2017}. Table \ref{Stab:model_inventory} lists the models used for each temperature target. \textbf{Middle Column:} Global sea-level rise (cm; relative to 2000) from the methodology of \citet{Kopp2014} (K14) (blue), using CMIP5 temperature trajectories from Left Column, and a semi-empirical global sea-level (SESL) model from \citet{Kopp2016} (red). Temperature trajectories that drive the SESL model are also shown in the Left Column. The thick line is the 50th percentile, heavy shading is the 17/83rd percentile, and light shading is the 5/95th percentile. \textbf{Right Column:} Probability distributions of projected 2050, 2100, and 2200 GMSL rise for GMST stabilization targets using the \citet{Kopp2014} framework (blue = 1.5 $^\circ$C, green = 2.0 $^\circ$C, orange = 2.5 $^\circ$C).}
\label{fig:gmst}
\end{figure}

The GMSL projections for each GMST target from the K14 and SESL method are shown in Fig. \ref{fig:gmst} and are tabulated along with the component contributions in Table \ref{tab:GMSLProjections}. For the K14 method, differences in median GMSL between 1.5 $^\circ$C, 2.0 $^\circ$C, and 2.5 $^\circ$C GMST stabilization targets do not appear until after 2050, when the 1.5 $^\circ$C scenario begins to separate from the 2.0 $^\circ$C and 2.5 $^\circ$C trajectories (Table \ref{tab:GMSLProjections}). The median GMST trajectories diverge earlier, around 2030 (Fig. \ref{Sfig:sesl_gmst}). This is consistent with the early to mid-century divergence in the radiative forcing pathways and this study's allocation of RCPs in the 1.5 $^\circ$C (primarily RCP2.6), 2.0 $^\circ$C (primarily RCP4.5), and 2.5 $^\circ$C (primarily RCP4.5 and RCP6) scenarios (Table \ref{Stab:model_inventory}). Median projections for 2100 GMSL under a 1.5 $^\circ$C scenario are 47 cm, with a \emph{likely} range (67\% probability) of 35--63 cm. An additional 8--11 cm of median GMSL rise is found for the 2.0 $^\circ$C and 2.5 $^\circ$C GMST scenarios, 55 cm (\emph{likely} 40--74 cm) and 58 cm (\emph{likely} 44--75 cm), respectively. The sources of GMSL projection variance are shown in SI Fig. \ref{Sfig:GSLvar}. Using the same framework, \citet{Kopp2014} found similar median 2100 GMSL projections under RCP2.6 and RCP4.5: 50 cm (\emph{likely} 37--65 cm) and 59 cm (\emph{likely} 45--77 cm), respectively.

Despite being warmer by a half-degree, the 2.5 $^\circ$C scenario largely overlaps the GMSL probability distribution for the 2.0 $^\circ$C scenario (Fig. \ref{fig:gmst}). For all GMST scenarios considered, TE contributes the most to median 2100 GMSL, and we find weak correlation ($r^{2}$ = 0.10) between 2100 GMST and the corresponding TE contribution (Fig. \ref{Sfig:zostoga_gmt}). The latter is due in part to variation in the transient climate response (TCR) and ocean heat uptake efficiency across CMIP5 models \citep{Kuhlbrodt2012,Raper2002}. To test the sensitivity of model-RCP filtering to the choice of GMST stabilization, we additionally calculate GMSL under a 1.75 $^\circ$C and 2.25 $^\circ$C scenario. The median 2100 GMSL under the 1.75 $^\circ$C scenario is 6 cm greater than the 1.5 $^\circ$C scenario, and the 2.25 $^\circ$C scenario is 1 cm less than the 2.0 $^\circ$C scenario (Table \ref{Stab:sensitivity}), suggesting that GMST scenarios that are primarily represented by only one RCP may be more sensitive to model filtering.

Agreement between central estimates from process-based and semi-empirical projections implies consistency with the observed statistical relationship between GMST and the rate of SLR used to calibrate the SESL model. Across scenarios, median 2100 GMSL projections from the SESL model driven with CMIP5 GMST trajectories are 6--7 cm lower than those from the K14 framework (Fig. \ref{fig:gmst} and Table \ref{tab:GMSLProjections}), a difference that is less in magnitude to those between the K14 projections and \citet{Kopp2016} SESL projections for RCPs 2.6 and 4.5 (\citealp{Kopp2016}, Table 2, showing median SESL projections of 38 cm and 51 cm for RCP2.6 and 4.5, vs. 50 and 59 cm for K14). The agreement between the processed-based and SESL projections is less when driven with the MAGICC GMST trajectories shown in SI Fig. \ref{Sfig:sesl_gmst} (median projection differences of 8--11 cm; Table \ref{tab:GMSLProjections}). Similarly, median estimates of 2100 GMSL for 1.5 $^\circ$C and 2.0 $^\circ$C scenarios from \citet{Schleussner2016} are 5--6 cm less than the projections using the K14 framework (Table \ref{tab:GMSLProjections}). The SLR projections of \citet{Schleussner2016} are based on a method that scales SLR component contributions as a function of GMST and ocean heat uptake \citep{Perrette2013}.

\subsection{Population inundation}
Under the median projected GMSL for a 2.0 $^\circ$C GMST stabilization, areas currently home to about 60 million people are at risk of being permanently submerged by 2150, including areas currently home to half a million inhabitants of United Nations defined Small Island Developing States (SIDS). By comparison, under the median projection for the 1.5 $^\circ$C stabilization scenario, areas currently home to about 5 million people, including 40,000 in SID avoid inundation (Table \ref{tab:inundation}). Aggregation of the SIDS can mask important risks. For instance, local SLR projections for 2150 under a 1.5 $^\circ$C GMST stabilization place areas currently home to almost a quarter of the current population of the Marshall Islands at risk of being permanently submerged. 

\subsection{Amplification of flood events}

\begin{figure}[h]
\centering
    \includegraphics[width=0.99\textwidth]{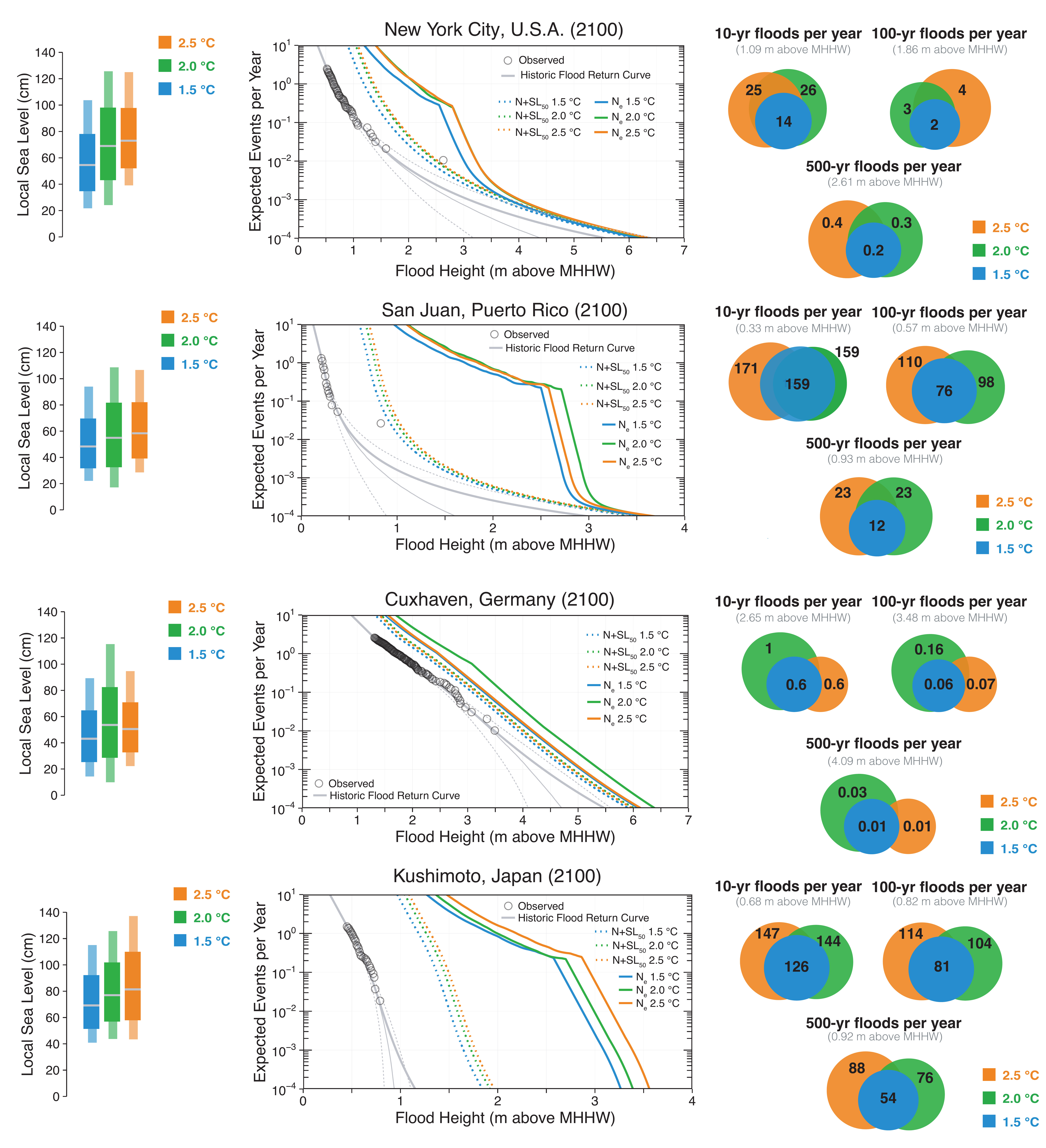} 
    \caption{\textbf{Top Left:} 2100 local sea-level rise (cm; relative to 2000) for New York City, U.S.A. under 1.5 $^\circ$C (blue), 2.0 $^\circ$C (green), and 2.5 $^\circ$C (orange) global mean surface temperature (GMST) stabilization. Grey bars are median, heavy colors are 17/83 percentile and light shading is 5/95 percentile. \textbf{Top Middle:} Flood return curves for New York City indicating the relationship between the number of expected flood events \emph{N}(\emph{z}) and flood level (\emph{z}) for different GMST stabilizations (blue = 1.5 $^\circ$C, green = 2.0 $^\circ$C, orange = 2.5$^\circ$C) and SLR assumptions for the year 2100. \emph{N} denotes the historic flood return curve (heavy grey line), grey circles are historical flood events, thin grey lines are the 17/50/83 percentiles of the GPD parameter uncertainty range, respectively. Median SLR in 2100 for each GMST stabilization are depicted by $N$ + SL$_{50}$ curves, and the expected flood return levels are depicted as $N_{e}$. \textbf{Top Right:} The expected flood amplification factor (AF) for New York City for 10, 100, and 500-yr flood events for 2100 under a 1.5 $^\circ$C (blue), 2.0 $^\circ$C (green), and 2.5 $^\circ$C (orange) GMST stabilization. \textbf{Second Row:} As for Top Row, but for San Juan, Puerto Rico. \textbf{Third Row:} As for Top Row, but for Cruxhaven, Germany. \textbf{Fourth Row:} As for Top Row, but for Kushimoto, Japan.}
\label{fig:returncurve}
\end{figure}

We assess the effects of different GMST stabilizations on coastal flooding by highlighting four cities: 1) New York, New York, USA 2) San Juan, Puerto Rico, USA 3) Cruxhaven, Lower Saxony, Germany, and 4) Kushimoto, Wakayama, Japan (Fig. \ref{fig:returncurve}). Estimates of the current 10-, 100-, and 500-yr flood (10\%, 1\%, 0.2\% probability per year) and the future flood amplification factor (AF) for all sites are provided in SI Tables S-4 to S-6. Under a 2.0 $^\circ$C GMST stabilization, the 2100 median local SLR for New York City is 69 cm, relative to 2000 (\emph{likely} 43--98 cm). In Fig. \ref{fig:returncurve}, median local SLR shifts the expected historic flood return curve to the right [i.e., N($z$), the heavy grey curve, becomes N+SL$_{50}$ 2.0 $^\circ$C, the dashed green curve] and increases the expected annual number of current 10-yr floods (a flood with a height of 1.09 m above MHHW) from 0.1/year to $\sim$10/year. However, when both the uncertainty in the GPD fit and the SLR projections are considered in the calculation of the projected future flood return curve (i.e., N$_{e}$ 2.0 $^\circ$C; the heavy green curve), the 10-yr flood amplification increases to 26/year (i.e., > 2/month). The discontinuities in the flood return curves demarcate the threshold where events are modeled with either a Gumbel distribution (from $\lambda$ to 182.6 events per year) and with the GPD \citep{Buchanan2016}. GHG mitigation that stabilizes GMST at 1.5 $^\circ$C reduces median local SLR at New York City to 55 cm (\emph{likely} 35--78 cm), and reduces the number of expected annual 10-yr flood events by half (13/year). By 2150, the reduction in expected 10-yr flood events from the 2.0 $^\circ$C to the 1.5 $^\circ$C scenario is still roughly 50\% (58/year vs. 97/year; Table \ref{Stab:af_table_rp10}).

Sea-level rise will increase the frequency of all flood events, but some flood events will amplify more than others. While higher frequency events (i.e., the current 10-yr flood event) are expected to increase the most for New York City by 2100, San Juan is expected to experience greater increases in lower frequency floods (i.e., 500-yr events). Specifically, GMST stabilized at 2.0 $^\circ$C is anticipated to produce 23 current 500-yr floods per year, on average (0.93 m above MHHW). If GMST warming is stabilized around 1.5 $^\circ$C, the expected number of 500-yr floods is reduced by roughly half (12/year, on average). At higher levels of SLR, the most frequent flood events (e.g., the 10-yr flood) become driven by tidal events, as opposed to storm surges. By 2100, the 10-year flood is expected to occur almost every other day under all scenarios. For some sites, the AF for the 2.0 $^\circ$C scenario may be greater than or equal to the AF for the 2.5 $^\circ$C scenario. This can be observed for projections around the Baltic and North Sea where there is large uncertainty in the sign of the ocean dynamics contribution to local sea-level change between models. After 2100, the GMSL overlap between the 2.0 $^\circ$C and 2.5 $^\circ$C scenario projections leads to indistinguishable differences in flood projections for some locations (Fig. \ref{fig:gmst} and Table \ref{tab:GMSLProjections})

In Asia, under a 2.0 $^\circ$C and 1.5 $^\circ$C GMST stabilization, by 2100, Kushimoto is projected to have median local SLR of 77 cm (\emph{likely} 57--102 cm) and 69 cm (\emph{likely} 51--92 cm), respectively; increasing the current number of expected 100-yr floods for Kushimoto from 1/100 years, on average, to 104/year and 81/year, on average. Some locations are projected to have less local SLR compared to New York, San Juan, and Kushimoto, and therefore less flood amplification. By 2100, under both a 2.0 $^\circ$C and 1.5 $^\circ$C GMST stabilization, Cruxhaven is projected to have median local SLR  increases of 54 cm (\emph{likely} 29--83 cm) and 43 cm (\emph{likely} 25--65 cm), respectively. Considering the entire projected probability distribution of SLR at Cruxhaven, the expected frequency of the current 500-yr flood event will become the future 100-yr flood. 

\begin{figure}[h]
\centering
    \includegraphics[width=0.99\textwidth]{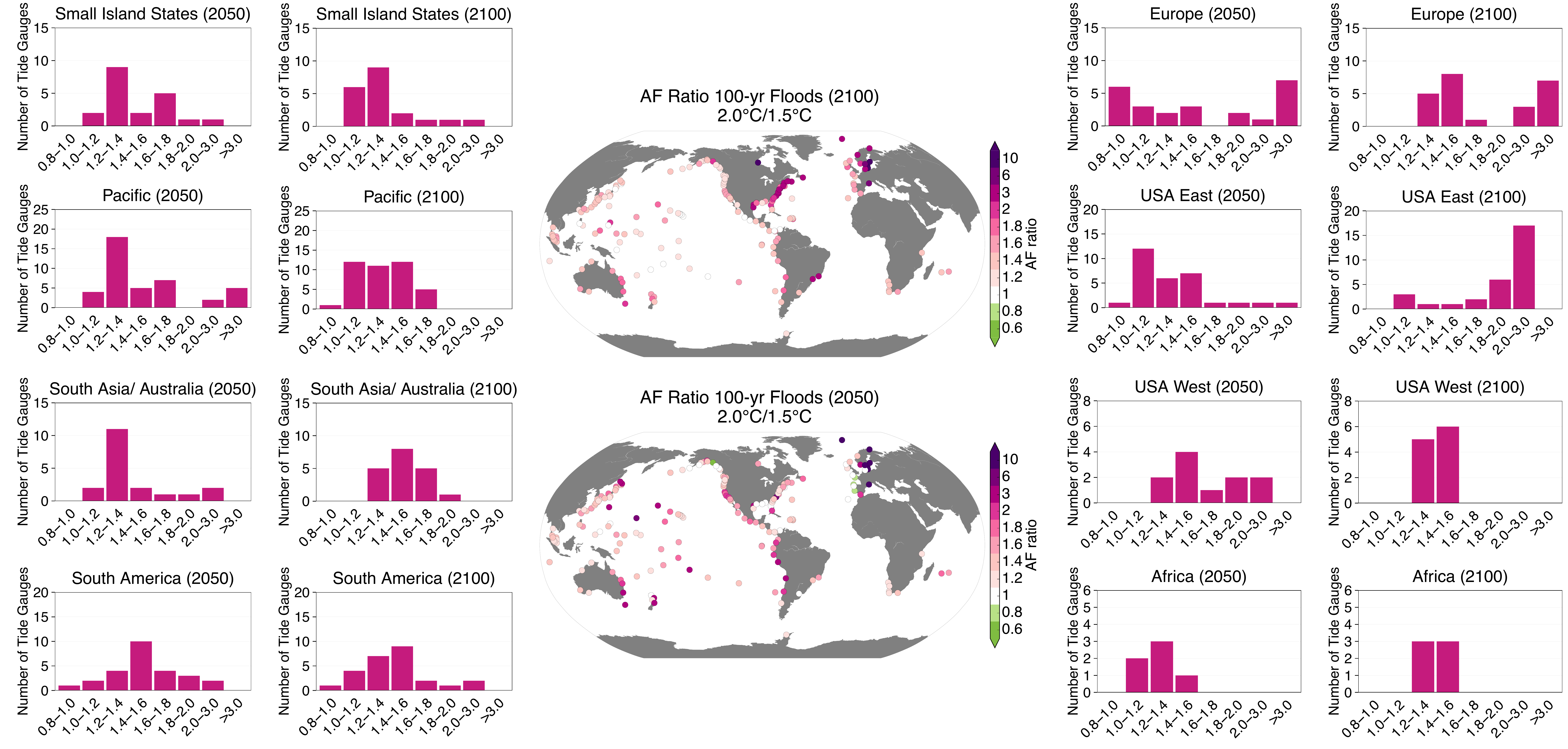} 
    \caption{\textbf{Maps:} The ratio of expected flood amplification factors (AFs) for 100-yr flood events between a 1.5 $^\circ$C and 2.0 $^\circ$C global mean surface temperature (GMST) stabilization target for the years 2050 and 2100. Larger 2.0 $^\circ$C/1.5 $^\circ$C AF ratios indicate locations where coastal flood benefits are greater from 1.5 $^\circ$C GMST stabilization, relative to a 2.0 $^\circ$C GMST stabilization. \textbf{Histograms:} Binned ratios of 2.0 $^\circ$C/1.5 $^\circ$C expected AFs for the 100-yr flood event for 2050 and 2100. ``Small Island States'' are Small Island Developing States defined by the United Nations. The list of sites included in each region are given in Table \ref{Stab:tgauges}.} \label{fig:af_maps} 
\end{figure}

We assess regional differences in 100-yr flood amplification between 2.0 $^\circ$C and 1.5 $^\circ$C GMST stabilization by binning ratios of 2.0 $^\circ$C/1.5 $^\circ$C expected AFs for 2050 and 2100 (Fig. \ref{fig:af_maps}). Bins on the right side of each graph become filled when there are flood benefits at stations from 1.5 $^\circ$C over 2.0 $^\circ$C GMST stabilization, while bins on the left side of each graph become filled when there are little or no benefits at stations from 1.5 $^\circ$C GMST over 2.0 $^\circ$C GMST stabilization. At mid-century, only a few sites indicate benefits from a 1.5 $^\circ$C GMST stabilization that are greater than 50\% reductions in 100-yr flood frequency as GMSL trajectories between scenarios have not appreciably separated from one another (Table \ref{tab:GMSLProjections}). However, by 2100, larger flood benefits of 1.5 $^\circ$C GMST stabilization are expected in Europe and the East and Gulf Coasts of the United States (U.S.), where flood amplification is reduced by roughly half. We find minimal flood benefits from achieving a 1.5 $^\circ$C GMST stabilization rather than a 2.0 $^\circ$C GMST stabilization for the West Coast of the U.S., the Pacific, and Indian Ocean regions (Fig. \ref{fig:af_maps}).

\section{Discussion and Conclusions}

The Paris Agreement seeks to stabilize GMST by limiting warming to ``well below 2.0 $^\circ$C above pre-industrial levels'', but a recent literature review under the UNFCCC found the notion that ``up to 2.0 $^\circ$C of warming is considered safe, is inadequate'' and that ``limiting global warming to below 1.5 $^\circ$C would come with several advantages'' \citep{UNFCCC2015b}. However, given the geographic diversity of climate impacts from any GMST target, there cannot be an objective threshold that defines when all impacts reach unmanageable levels. The location-specific increases in the frequency of coastal floods is one such example. The selection of a GMST target has important implications for long-term GMSL rise and, consequently, coastal flooding. Assessing the distribution of impacts of incremental levels of warming on coastal flooding is of relevance to > 625 million people who currently reside in low-lying coastal areas \citep{Neumann2015} and are vulnerable to current and future flood events. For countries without the economic and physical capacity to construct flood protection and flood-resilient infrastructure\textemdash including some recognized by the United Nations as Small Island Developing States\textemdash local SLR that results in permanent inundation and unmanageable flooding may threaten their existence \citep{Wong2014,Diaz2016}. The only feasible option for maintaining habitability for these locations may be the management of GMST through international climate accords, like the Paris Agreement, that govern the long-term committed rise in GMSL.

Only considering changes to the mean local sea level, we find that roughly 5 million fewer inhabitants currently reside in lands that will be permanently submerged by 2150 under a 1.5 $^\circ$C GMST stabilization compared to that of the 2.0 $^\circ$C case, including 40,000 fewer inhabitants of SIDS (Table \ref{tab:inundation}). The effects of GMST stabilization on coastal flooding varies greatly by region and by return level (e.g., the 10-yr versus the 100-yr flood, ect.). Globally, for the current 100-yr flood, we find that by 2100, the Eastern and Gulf coasts of the U.S. and Europe could benefit the most from a 1.5 $^\circ$C GMST stabilization relative to a 2.0 $^\circ$C GMST stabilization, with flood frequency amplification being reduced by about half. However, while fractional reductions may appear substantial in some cases, small absolute differences may warrant similar coastal flood risk management responses. For instance, for New York City, we estimate that the difference in the expected number of current 100-yr floods per year (0.01 events per year, on average) between a 2.0 $^\circ$C to a 1.5 $^\circ$C GMST stabilization is only 3 times per year to 2 times per year, on average (Fig. \ref{fig:returncurve}). 

While these data could be used in support of local probabilistic risk management strategies that intend to reduce current and future exposure and vulnerability to extreme flood events, some caveats should be highlighted. First, while our projections carry probabilities, the probabilities should be viewed in light of the characterization of the uncertainty of each sea level component. The true probability distribution of some sea level components may be imperfectly sampled. Second, changes to storm frequency and severity as well as effects from waves could significantly influence future flood events \citep[e.g.,][]{Reed2015,Hatzikyriakou2017}. In this study, wave effects are not considered, and we assume that the frequency of storm arrivals and their intensity will remain constant\textemdash and thus the Poisson and GPD scale and shape parameters. Modifications could be made to encompass changes in these parameters with time \citep{Emanuel2013,Knutson2010}. Lastly, these projections are for specific tide gauge locations and they may not be representative of the greater vicinity in which they are located. 

The selection of the level at which to stabilize the GMST in the coming years will determine the committed amounts of future GMSL \citep{Clark2016,Levermann2013}. Our projected coastal flood impacts through the end of the 22nd century should be placed in the context of longer timeframes. Stabilization of GMST does not imply stabilization of GMSL. Regardless of the mitigation scenario chosen, GMSL rise due to TE is expected to continue for centuries to millennia. Additionally, some studies suggest that sustained GMST warming above given thresholds, including those as low as 2 $^\circ$C, will lead to a near-complete loss of the GIS over a millennium or more \citep[][section 13.4.3]{Church2013}. Coincident with continued GMSL rise will be further increases in the frequency of current flood events and an increasing number of currently inhabited areas that will be permanently submerged. A comprehensive approach to managing coastal flood risks would take into account changes on these very long time frames.

\begin{acknowledgements}
We thank \hl{XX} anonymous reviewers and Carl-Friedrich Schleussner and colleagues at Climate Analytics (Berlin) for the helpful discussions. REK was supported in part by a grant from Rhodium Group (for whom he has previously worked as a consultant), as part of the Climate Impact Lab consortium and in part NSF grant ICER-1663807. We acknowledge the World Climate Research Programme's Working Group on Coupled Modeling, which is responsible for CMIP, and we thank the climate modeling groups (listed in Supporting Information Table \ref{Stab:model_inventory}) for producing and making available their model output. For CMIP, the U.S. Department of Energy's Program for Climate Model Diagnosis and Intercomparison provides coordinating support and led development of software infrastructure in partnership with the Global Organization for Earth System Science Portals. Code for generating sea-level projections is available in the ProjectSL (https://github.com/bobkopp/ProjectSL), LocalizeSL (https://github.com/bobkopp/LocalizeSL), and SESL (https://github.com/bobkopp/SESL) repositories on Github. Code for generating flood projections is available in the \hl{XXXX (D.J. will upload)} repository on Github. The statements, findings, conclusions, and recommendations are those of the authors and do not necessarily reflect the views of the funding agencies.

\end{acknowledgements}

\bibliographystyle{spbasic_etal}
\bibliography{ipcc_sealevel}

\newpage

\section{Figures}

\begin{table}[h]
\centering
\setlength{\tabcolsep}{4pt}
\caption{GMSL  projections. All values are cm above 2000 CE baseline. AIS = Antarctic Ice Sheet, GIS = Greenland Ice Sheet; TE = Thermal Expansion; GIC = Glacial Ice Melt; LWS = Land-Water Storage. K16: Semi-empirical sea level (SESL) model from \citet{Kopp2016} driven with global mean surface temperature (GMST) trajectories from MAGICC (see SI Fig. \ref{Sfig:sesl_gmst}) and CMIP5 GMST trajectories (see Fig. \ref{fig:gmst}); J16: \citet{Jevrejeva2016}; S16: \citet{Schleussner2016}; S12: \citet{Schaeffer2012}. $^{*}$the estimate for \citet{Jevrejeva2016} is not from 2100, rather it is upon reaching a median 2.0 $^\circ$C GMST increase at mid-century in an RCP8.5 ensemble. }
{\small
\begin{tabular}{l|ccc|ccc|ccc|}
&\multicolumn{3}{c|}{1.5 $^\circ$C}&\multicolumn{3}{c|}{2.0 $^\circ$C}&\multicolumn{3}{c|}{2.5 $^\circ$C}  \\
cm&50   &  17--83   &  5--95  &50   &  17--83   &  5--95  &50   &  17--83   &  5--95  \\
\hline  \multicolumn{10}{l}{2100---Components}  \\
AIS&6&-4--17&-8--35&6&-4--17&-8--35&5&-5--16&-9--33  \\
GIS&6&4--12&3--17&6&4--12&3--17&9&4--15&2--23  \\
TE&19&14--23&10--27&25&16--34&9--42&26&20--31&16--35  \\
GIC&11&8--13&6--15&12&7--16&4--21&13&11--15&9--17  \\
LWS&5&3--7&2--8&5&3--7&2--8&5&3--7&2--8  \\
\hline
Total&47&35--64&28--82&55&40--75&30--94&58&44--75&36--93  \\
\hline  \multicolumn{10}{l}{Projections  by  year}  \\
2050&24&20--28&18--32&25&20--32&17--37&26&21--30&19--34  \\
2070&33&27--41&23--49&37&28--48&23--58&38&31--47&27--55  \\
2100&47&35--64&28--82&55&40--75&30--94&58&44--75&36--93  \\
2150&68&41--106&28--150&89&54--134&35--178&86&53--128&35--169  \\
2200&91&41--159&19--240&120&64--197&32--277&117&61--192&30--269  \\

\hline  \multicolumn{10}{l}{Other  projections  for  2100}  \\
K16$^{1}$ &38&33--43&30--47&45&39--52&35--58&54&47--62&42--68  \\
K16$^{2}$ &41&36--48&32--53&48&41--56&36--62&51&45--59&41--65  \\
J16$^{*}$ &--&--&--&22&--&15--33&--&--&-- \\
S16&41&29--53&--&50&36--65&--&--&--&-- \\
S12 &77&--&54--99&80&--&56--105&--&--&--\\
\hline  \multicolumn{10}{l}{Other  projections  for  2200}  \\
S12 &135&--&85--195&180&--&110--345&--&--&--\\
\hline
\multicolumn{10}{l}{$^{1}$ SESL model driven with MAGICC6 GMST trajectories shown in SI Fig. \ref{Sfig:sesl_gmst}} \\
\multicolumn{10}{l}{$^{2}$ SESL model driven with CMIP5 GMST trajectories shown in Fig. \ref{fig:gmst}} 
\end{tabular}}
\label{tab:GMSLProjections}
\end{table}

\newpage
\clearpage

\begin{table}[h]
\centering
    \caption{Human population (in millions) currently residing on lands at risk of permanent inundation based on median (5--95th percentile) local SLR projections. Population estimates are from 2010. The top five countries with the most exposure in 2150 are included in the table as well as United Nations defined Small Island Developing States (SIDS).}
\begin{tabular}{ccccc}
    \multicolumn{5}{l}{Human population exposure under 2100 local SLR projections (millions)} \\
    
    \hline
    Region & Total Pop. & 1.5 $^\circ$C & 2.0 $^\circ$C & 2.5 $^\circ$C \\
   \hline
    World & 6,836.42 & 45.88 (31.87--68.83) & 48.23 (31.99--78.38) & 50.27 (33.15--77.28) \\
    China & 1,330.20 & 11.59 (5.87--20.22) & 12.53 (5.98--21.69) & 13.25 (6.12--22.93) \\
    Vietnam & 89.55 & 6.55 (4.55--9.85) & 6.89 (4.58--10.44) & 7.14 (4.65--11.07) \\
    Japan & 126.66 & 4.43 (3.82--5.54) & 4.59 (3.87--5.77) & 4.69 (3.88--6.10) \\
    Netherlands & 16.78 & 4.71 (4.19--5.56) & 4.86 (4.18--5.88) & 4.84 (4.35--5.63) \\
    Bangladesh & 156.13 & 2.81 (1.98--4.29) & 3.00 (2.06--4.66) & 3.09 (2.12--4.92) \\
    SIDS  & 62.08 & 0.40 (0.30--0.55) & 0.42 (0.30--0.63) & 0.43 (0.31--0.63) \\
    \hline
    \end{tabular}  \label{tab:addlabel}%

    \begin{tabular}{ccccc}
    \multicolumn{5}{l}{Human population exposure under 2150 local SLR projections (millions)} \\
    \hline
    Region & Total Pop. & 1.5 $^\circ$C & 2.0 $^\circ$C & 2.5 $^\circ$C \\
   \hline
    World & 6,836.42 & 55.49 (32.45--111.58) & 60.48 (32.83--133.88) & 61.95 (33.72--127.07) \\
    China & 1,330.20 & 14.22 (5.72--30.54) & 16.38 (5.84--35.28) & 16.51 (5.69--36.73) \\
    Vietnam & 89.55 & 7.53 (4.46--14.99) & 8.3 (4.50--16.61) & 8.27 (4.52--16.62) \\
    Japan & 126.66 & 4.89 (3.82--5.54) & 5.3 (3.87--5.77) & 5.34 (3.88--6.10) \\
    Netherlands & 16.78 & 5.06 (4.19--5.56) & 5.17 (4.18--5.88) & 5.26 (4.35--5.63) \\
    Bangladesh & 156.13 & 4.43 (1.98--4.29) & 4.98 (2.06--4.66) & 4.98 (2.12--4.92) \\
    SIDS  & 62.08 & 0.46 (0.28--0.90) & 0.50 (0.29--1.11) & 0.51 (0.30--1.01) \\
    \hline
    \end{tabular}  \label{tab:inundation}%
\end{table}%

\newpage
\clearpage

\iftoggle{supplement}{\startsupplement

\section*{Supplementary Information}

\section{Preparation of tide gauge data for extreme value analysis}
Tide gauge observations are prepared as the basis for extreme value analysis following the methods of \citet{Tebaldi2012}. First, daily maximum tide gauge values are calculated from the ``Research Quality'' hourly observations from the University of Hawaii Sea Level Center \citep[retrieved from uhslc.soest.hawaii.edu, June 2017; ][]{Caldwell2015}. Only tide gauge locations with record lengths $\ge$ 30 years and $\ge$ 80 percent data completion are considered. A list of tide gauges and their record lengths is provided in the Table \ref{Stab:tgauges}. The impact of day-to-day weather, astronomical tides and seasonal cycles on sea level is isolated by removing sea level change over the tide gauge record. Monthly-mean sea levels over the tide gauge record are used to linearly de-trend the daily maximum observations. Mean higher high water (MHHW) at each tide gauge is estimated using the average of the daily maximum tide gauge observations over the most recent 19-year period. While the MHHW calculation approach differs from the current U.S. standard (which is defined over the National Tidal Datum Epoch of 1983--2001), it is taken after de-trending of the time series and therefore should be close to stationary. Each de-trended tide gauge series is then referenced to its own MHHW level. Finally, at each tide gauge the daily observations above the 99th percentile are de-clustered to separate multiple observations made during the same extreme flood event and so that each observation is independent of one another. The 99th percentile is used as it is generally above the highest seasonal tide and it balances the bias-variance trade-off in the GPD parameter estimation. If too low of a GPD threshold is chosen, more observations than those exclusively in the tail of the GPD distribution might end up being included in the parameter calculation, causing bias. If too high of a GPD threshold is chosen, then too few observations may be incorporated in the estimation of distribution parameters leading to greater variance, relative to a case that uses more observations.

\begin{figure}[H]
\centering
    \includegraphics[width=0.75\textwidth]{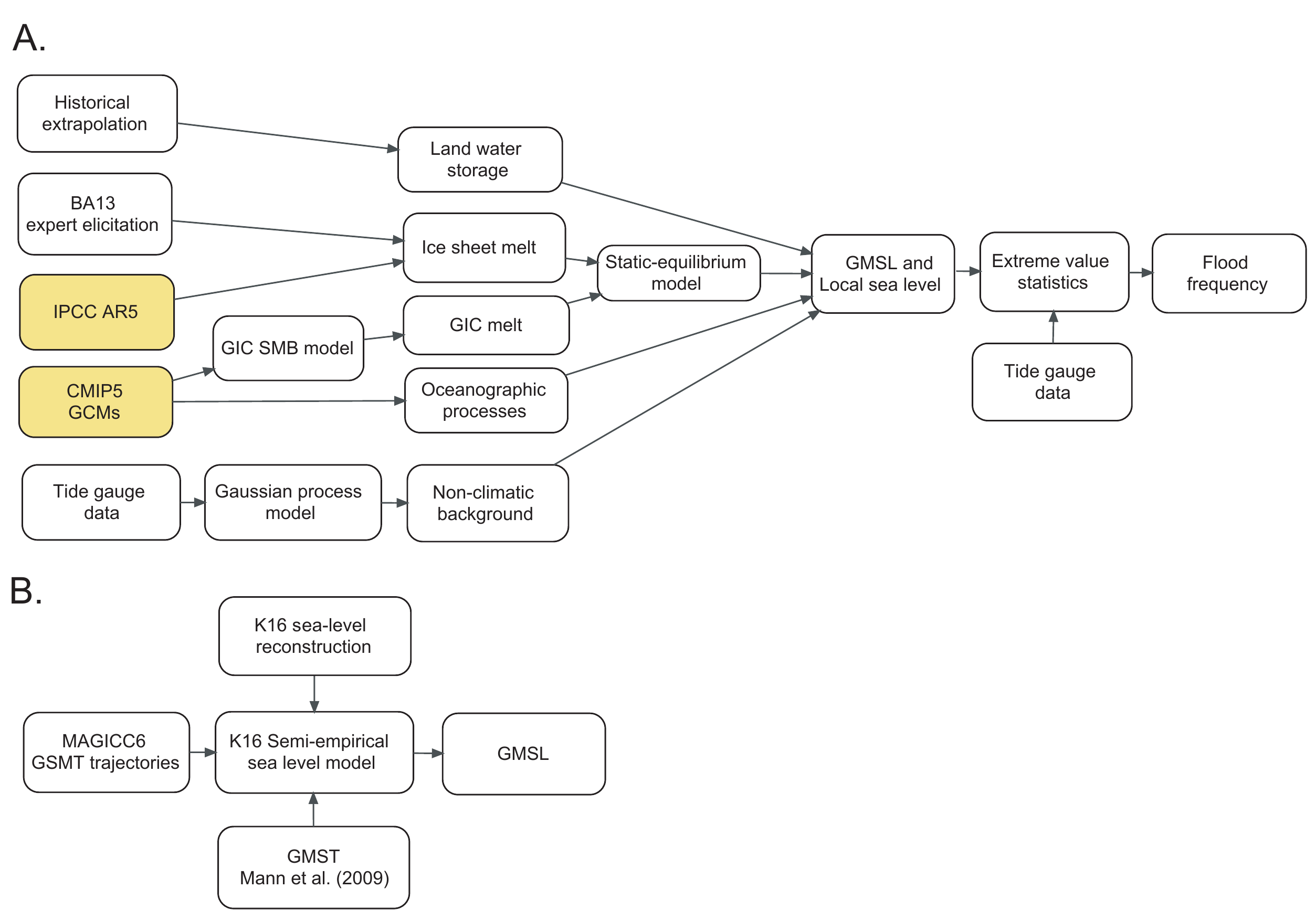} 
\caption{\textbf{Top:} Logical flow of sources of information used in local sea-level projections and flood frequency return curves. GCMs are global climate models; GIC is glacial ice
contribution; SMB is surface mass balance; GMT is global mean temperature; GMSL is global mean sea level; BA13 is \citealp{Bamber2013}; K16 is \citealp{Kopp2016}. Orange shades indicate where RCP and model grouping occurs (see Table \ref{Stab:model_inventory}). \textbf{Bottom:} Logical flow of sources of information used to construct semi-empirical sea level GMSL projections.}
\label{Sfig:flow_diag}
\end{figure}

\newpage 

\begin{figure}[H]
\centering
    \includegraphics[width=0.45\textwidth]{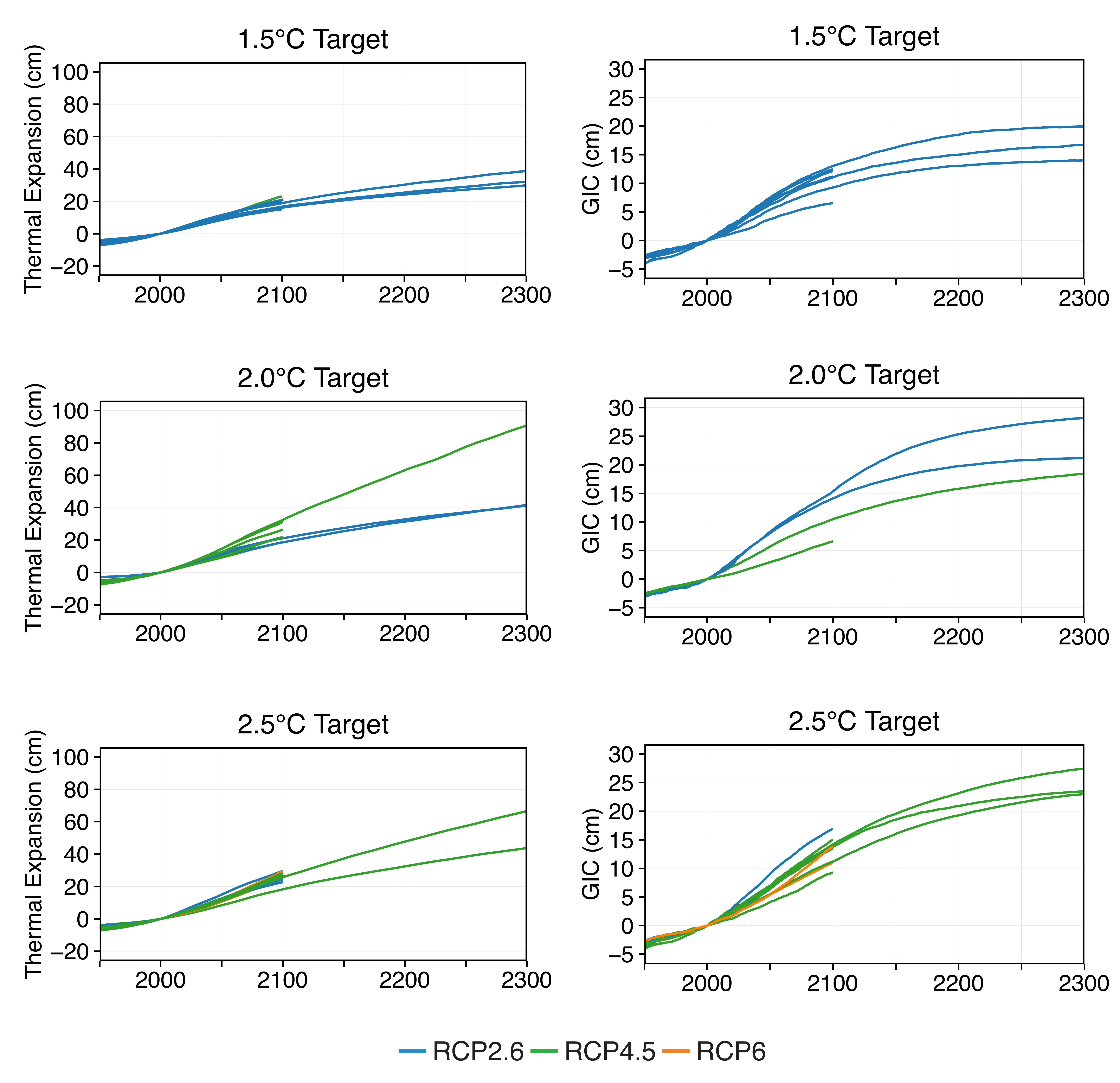} 
\caption{\textbf{Left Column:} Thermal expansion contribution to global mean sea-level (GMSL) rise (cm; relative to 2000) from CMIP5 models that have been smoothed and corrected for model drift for global mean surface temperature stabilization targets of 1.5 $^\circ$C, 2.0 $^\circ$C, and 2.5 $^\circ$C (blue = RCP2.6, green = RCP4.5, orange = RCP6). \textbf{Right Column:} As for Left Column, but for the glacial ice contribution (GIC) to GMSL rise (cm; relative to 2000) using the model from \citet{Marzeion2012}. Table \ref{Stab:model_inventory} lists the models used for each temperature target.} 
\label{Sfig:zostoga_gic}
\end{figure}

\newpage

\begin{figure}[H]
\centering
    \includegraphics[width=0.75\textwidth]{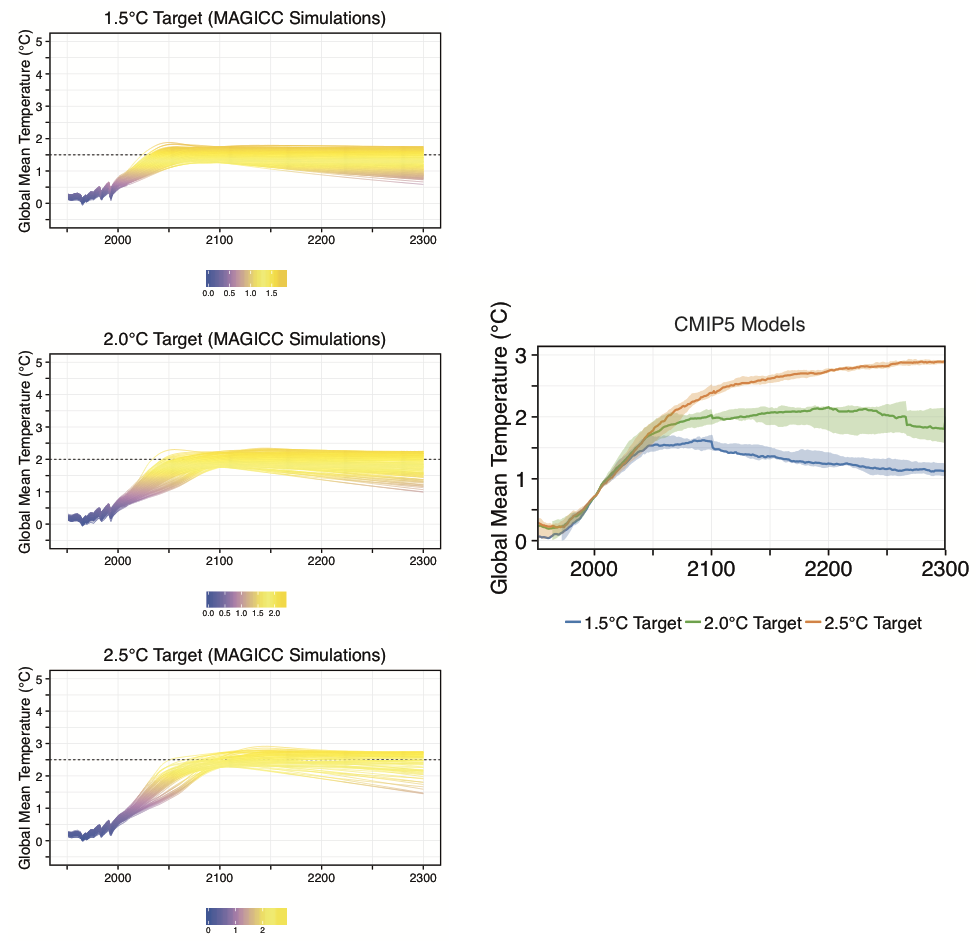} 
\caption{\textbf{Left:}  Global mean temperature trajectories from MAGICC6 for 1.5$^\circ$C, 2.0$^\circ$C, and 2.5$^\circ$C temperature targets at 2100. Temperatures are relative to 1875--1900. \textbf{Right:} Global mean surface temperature (GMST) trajectories from CMIP5 models (1950--2300) that have a 2100 GMST of 1.5 $^\circ$C, 2.0 $^\circ$C, and 2.5 $^\circ$C $\pm$ 0.25$^\circ$C (relative to 1875--1900). GMST is anomalized to 1991--2009 and shifted up by 0.72$^\circ$C to account for warming since 1875--1900 \citep{Hansen2010,GISS2017}. Solid line is the 50th percentile and light shading is the 17th/83rd range.}
\label{Sfig:sesl_gmst}
\end{figure}

\newpage 

\begin{figure}[H]
\centering
    \includegraphics[width=0.5\textwidth]{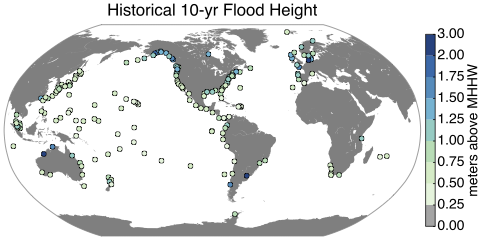} 
    \includegraphics[width=0.5\textwidth]{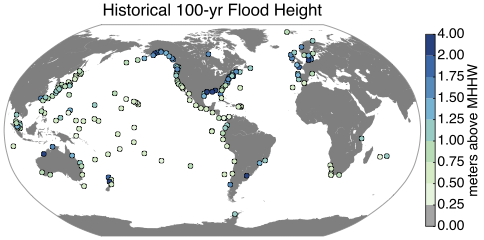} 
    \includegraphics[width=0.5\textwidth]{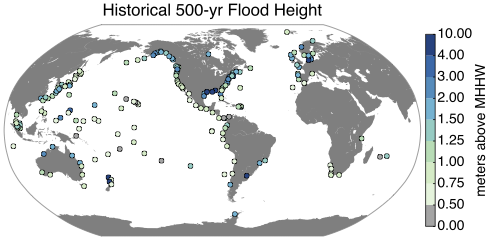}
\caption{Historical flood height [meters above mean higher high water (MHHW)] of floods with return periods of 10-, 100-, and 500-years.}
\label{Sfig:histFloodMaps}
\end{figure}

\newpage 

\begin{figure}[H]
\centering
    \includegraphics[width=0.75\textwidth]{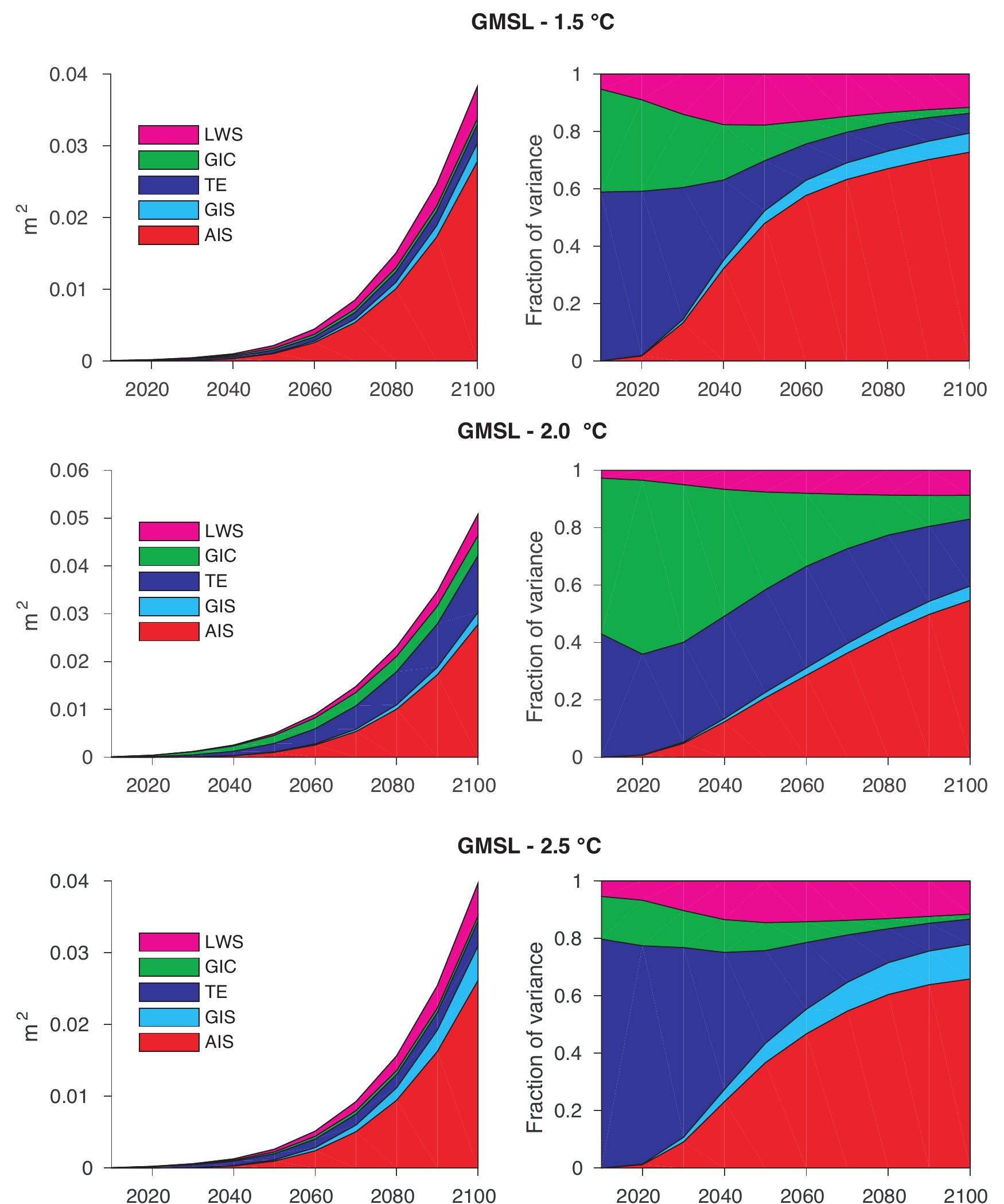} 
\caption{Global mean sea level (GMSL) sources of variance in raw and fractional terms in 1.5 $^\circ$C, 2.0 $^\circ$C, and 2.5 $^\circ$C global mean surface temperature stabilization scenarios. AIS: Antarctic ice sheet, GIS: Greenland ice sheet, TE: thermal expansion, GIC: glaciers and ice caps, LWS: land water storage}
\label{Sfig:GSLvar}
\end{figure}

\newpage 

\begin{figure}[H]
\centering
    \includegraphics[width=0.45\textwidth]{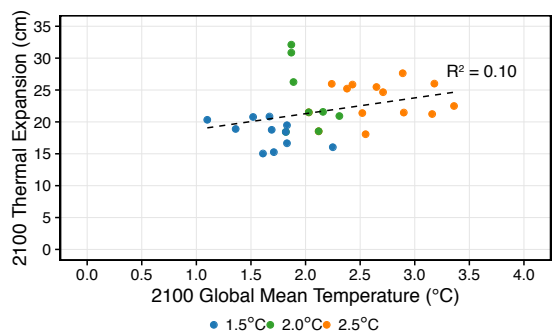} 
\caption{Relationship between 2100 global mean thermal expansion contribution to sea-level rise (i.e., 'zostoga') (cm) and the 19-yr running average of global mean surface temperature (GMST) for 2100 from CMIP5 model output ($^\circ$C, relative to 1875--1900) under 1.5 $^\circ$C (blue), 2.0 $^\circ$C (green), and 2.5 $^\circ$C (orange) GMST stabilization scenarios. Black dotted line is the linear fit across all temperature scenarios and all models.} 
\label{Sfig:zostoga_gmt}
\end{figure}

\newpage 

{\small
 \begin{longtable}{lllcccccc}
     \caption{List of tide gauges used from the University of Hawaii Sea Level Center and their record lengths.} \\
     Site  & Country & Region &  Lat & Lon  & UHawaii ID &  Start & End & Length (yrs)  \\
    \hline
\endfirsthead
\multicolumn{9}{c}
{{\bfseries \tablename\ \thetable{} -- continued from previous page}} \\
 Site  & Country & Region &  Lat & Lon  & UHawaii ID &  Start & End & Length (yrs)  \\
\hline
\endhead
\hline
\multicolumn{9}{c}{{Continued on next page}} \\
\endfoot
\endlastfoot 
  Buenos Aires & Argentina & South America & -34.67 & -58.50 & 285a  & 1905  & 1961  & 57 \\
    Fort Denison & Australia & South Asia/ Australia & -33.90 & 151.32 & 333a  & 1965  & 2015  & 51 \\
    Bundaberg & Australia & South Asia/ Australia & -24.77 & 152.50 & 332a  & 1984  & 2015  & 32 \\
    Brisbane & Australia & South Asia/ Australia & -27.37 & 153.17 & 331a  & 1984  & 2015  & 32 \\
    Spring Bay & Australia & South Asia/ Australia & -42.67 & 148.07 & 335a  & 1985  & 2015  & 31 \\
    Townsville & Australia & South Asia/ Australia & -19.25 & 146.83 & 334a  & 1984  & 2013  & 30 \\
    Broome & Australia & South Asia/ Australia & -18.02 & 122.23 & 166a  & 1986  & 2015  & 30 \\
    Cocos & Australia & South Asia/ Australia & -12.12 & 96.97 & 171a  & 1985  & 2015  & 31 \\
    Darwin & Australia & South Asia/ Australia & -12.52 & 130.97 & 168a  & 1984  & 2015  & 32 \\
    Esperance & Australia & South Asia/ Australia & -33.90 & 122.02 & 176a  & 1985  & 2015  & 31 \\
    Fremantle & Australia & South Asia/ Australia & -32.05 & 115.73 & 175a  & 1984  & 2015  & 32 \\
    Cananeia & Brazil & South America & -25.02 & -48.00 & 281a  & 1954  & 2006  & 53 \\
    Ilha Fiscal, RJ & Brazil & South America & -23.02 & -43.30 & 280a  & 1963  & 2012  & 50 \\
    Victoria, BC & Canada & Canada & 48.50 & -123.40 & 543a  & 1909  & 2014  & 106 \\
    Prince Rupert & Canada & Canada & 54.32 & -130.38 & 540a  & 1924  & 2014  & 91 \\
    Tofino & Canada & Canada & 49.18 & -126.03 & 542a  & 1930  & 2014  & 85 \\
    St. John's-A & Canada & Canada & 47.57 & -52.70 & 276a  & 1952  & 1989  & 38 \\
    Halifax & Canada & Canada & 44.67 & -63.58 & 275a  & 1899  & 2014  & 116 \\
    Churchill & Canada & Canada & 58.77 & -94.18 & 274a  & 1961  & 2012  & 52 \\
    Puerto Montt & Chile & South America & -41.50 & -73.07 & 684a  & 1980  & 2014  & 35 \\
    Juan Fernandez-B & Chile & South America & -33.67 & -78.95 & 021b  & 1985  & 2014  & 30 \\
    Antofagasta & Chile & South America & -23.68 & -70.45 & 080a  & 1945  & 2014  & 70 \\
    Easter-C & Chile & South America & -27.20 & -109.50 & 022c  & 1970  & 2014  & 45 \\
    Valparaiso & Chile & South America & -33.12 & -71.72 & 081a  & 1944  & 2014  & 71 \\
    Xiamen & China & Pacific & 24.45 & 118.07 & 376a  & 1954  & 1997  & 44 \\
    Buenaventura & Colombia & South America & 3.95  & -77.17 & 085a  & 1953  & 2014  & 62 \\
    Tumaco & Colombia & South America & 1.82  & -78.85 & 303a  & 1951  & 2014  & 64 \\
    Cartagena & Colombia & South America & 10.38 & -75.53 & 265a  & 1951  & 1993  & 43 \\
    Penrhyn & Cook Islands & SIDS  & -9.07 & -158.08 & 024a  & 1977  & 2015  & 39 \\
    Quepos-A & Costa Rica & South America & 9.40  & -84.17 & 087a  & 1961  & 1994  & 34 \\
    Hornbaek & Denmark & Europe & 56.10 & 12.47 & 838a  & 1891  & 2012  & 122 \\
    Gedser & Denmark & Europe & 54.57 & 11.93 & 837a  & 1891  & 2012  & 122 \\
    Baltra-B & Ecuador & South America & -0.47 & -90.30 & 003b  & 1985  & 2015  & 31 \\
    Santa Cruz & Ecuador & South America & -0.80 & -90.43 & 030a  & 1978  & 2015  & 38 \\
    La Libertad & Ecuador & South America & -2.20 & -80.92 & 091a  & 1949  & 2015  & 67 \\
    Acajutla-A & El Salvador & South America & 13.58 & -89.83 & 082a  & 1962  & 2001  & 40 \\
    Chuuk & Fd. St. Micronesia & SIDS  & 7.45  & 151.85 & 054a  & 1956  & 1991  & 36 \\
    Kapingamarangi & Fd. St. Micronesia & SIDS  & 1.23  & 154.87 & 029a  & 1978  & 2015  & 38 \\
    Pohnpei-B & Fd. St. Micronesia & SIDS  & 7.02  & 158.33 & 001b  & 1974  & 2004  & 31 \\
    Yap-B & Fd. St. Micronesia & SIDS  & 9.58  & 138.23 & 008b  & 1969  & 2015  & 47 \\
    Suva-C & Fiji  & SIDS  & -18.27 & 178.52 & 018c  & 1972  & 2015  & 44 \\
    Noumea & France & Europe & -22.32 & 166.42 & 019a  & 1967  & 2015  & 49 \\
    Brest & France & Europe & 48.38 & -4.60 & 822a  & 1846  & 2014  & 169 \\
    Marseille & France & Europe & 43.38 & 5.38  & 824a  & 1885  & 1988  & 104 \\
    Rikitea & French Polynesia & SIDS  & -23.20 & -134.98 & 016a  & 1969  & 2015  & 47 \\
    Papeete-B & French Polynesia & SIDS  & -17.60 & -149.57 & 015b  & 1975  & 2015  & 41 \\
    Cuxhaven & Germany & Europe & 53.87 & 8.72  & 825a  & 1917  & 2014  & 98 \\
    Malin Head & Ireland & Europe & 55.37 & -7.33 & 834a  & 1958  & 2001  & 44 \\
    Hakodate & Japan & Pacific & 41.78 & 140.72 & 364a  & 1964  & 2014  & 51 \\
    Hamada & Japan & Pacific & 34.90 & 132.07 & 348a  & 1984  & 2014  & 31 \\
    Maisaka & Japan & Pacific & 34.68 & 137.62 & 356a  & 1968  & 2014  & 47 \\
    Ishigaki & Japan & Pacific & 24.33 & 124.15 & 365a  & 1969  & 2014  & 46 \\
    Naha  & Japan & Pacific & 26.22 & 127.67 & 355a  & 1966  & 2014  & 49 \\
    Toyama & Japan & Pacific & 36.77 & 137.23 & 349a  & 1967  & 2014  & 48 \\
    Hosojima & Japan & Pacific & 32.42 & 131.68 & 358a  & 1933  & 1975  & 43 \\
    Kushiro & Japan & Pacific & 42.97 & 144.37 & 350a  & 1963  & 2014  & 52 \\
    Abashiri & Japan & Pacific & 44.02 & 144.28 & 347a  & 1968  & 2014  & 47 \\
    Mera  & Japan & Pacific & 34.92 & 139.82 & 352a  & 1965  & 2014  & 50 \\
    Wakkanai & Japan & Pacific & 45.40 & 141.68 & 360a  & 1967  & 2014  & 48 \\
    Chichijima & Japan & Pacific & 27.10 & 142.18 & 047a  & 1975  & 2014  & 40 \\
    Nishinoomote & Japan & Pacific & 30.75 & 131.07 & 363a  & 1965  & 2013  & 49 \\
    Naze  & Japan & Pacific & 28.52 & 129.60 & 359a  & 1957  & 2013  & 57 \\
    Hachinohe & Japan & Pacific & 40.53 & 141.53 & 375a  & 1980  & 2011  & 32 \\
    Miyakejima & Japan & Pacific & 34.07 & 139.62 & 357a  & 1964  & 2013  & 50 \\
    Nakano Shima & Japan & Pacific & 29.92 & 129.92 & 345a  & 1984  & 2013  & 30 \\
    Ofunato & Japan & Pacific & 39.02 & 141.75 & 351a  & 1965  & 2014  & 50 \\
    Nagasaki & Japan & Pacific & 32.73 & 129.87 & 362a  & 1985  & 2014  & 30 \\
    Aburatsu & Japan & Pacific & 31.58 & 131.42 & 354a  & 1961  & 2014  & 54 \\
    Kushimoto & Japan & Pacific & 33.48 & 135.77 & 353a  & 1961  & 2014  & 54 \\
    Cendering & Malaysia & South Asia/ Australia & 5.40  & 103.22 & 320a  & 1984  & 2013  & 30 \\
    Johor Baharu & Malaysia & South Asia/ Australia & 1.57  & 103.87 & 321a  & 1983  & 2013  & 31 \\
    Kuantan & Malaysia & South Asia/ Australia & 4.05  & 103.55 & 322a  & 1983  & 2013  & 31 \\
    Keling & Malaysia & South Asia/ Australia & 2.35  & 102.18 & 141a  & 1984  & 2013  & 30 \\
    Lumut & Malaysia & South Asia/ Australia & 4.30  & 100.73 & 143a  & 1984  & 2013  & 30 \\
    Kelang & Malaysia & South Asia/ Australia & 3.05  & 101.43 & 140a  & 1983  & 2013  & 31 \\
    Langkawi & Malaysia & South Asia/ Australia & 6.90  & 99.80 & 142a  & 1985  & 2015  & 31 \\
    Penang & Malaysia & South Asia/ Australia & 5.47  & 100.47 & 144a  & 1984  & 2013  & 30 \\
    Port Louis-C & Mauritius & SIDS  & -20.20 & 57.60 & 103c  & 1986  & 2015  & 30 \\
    Rodrigues & Mauritius & SIDS  & -19.68 & 63.43 & 105a  & 1986  & 2015  & 30 \\
    Manzanillo-A & Mexico & South America & 19.08 & -104.45 & 395a  & 1953  & 1982  & 30 \\
    Ensenada & Mexico & South America & 31.85 & -116.63 & 317a  & 1956  & 1991  & 36 \\
    Salina Cruz & Mexico & South America & 16.25 & -95.23 & 394a  & 1952  & 1983  & 32 \\
    Acapulco-A, Gro. & Mexico & South America & 16.90 & -100.02 & 316a  & 1952  & 1995  & 44 \\
    Cabo San Lucas & Mexico & South America & 23.00 & -109.98 & 034a  & 1973  & 2002  & 30 \\
    Guaymas & Mexico & South America & 27.93 & -110.90 & 397a  & 1953  & 1986  & 34 \\
    Saipan-B & N. Mariana Islands & SIDS  & 15.32 & 145.82 & 028b  & 1978  & 2015  & 38 \\
    Marsden Point & New Zealand & South Asia/ Australia & -35.83 & 174.50 & 398a  & 1975  & 2014  & 40 \\
    Tauranga & New Zealand & South Asia/ Australia & -37.65 & 176.18 & 073a  & 1984  & 2014  & 31 \\
    Taranaki & New Zealand & South Asia/ Australia & -39.05 & 174.03 & 076a  & 1984  & 2014  & 31 \\
    Wellington & New Zealand & South Asia/ Australia & -41.28 & 174.78 & 071a  & 1944  & 2014  & 71 \\
    Tregde & Norway & Europe & 58.00 & 7.57  & 804a  & 1927  & 2008  & 82 \\
    Rorvik & Norway & Europe & 64.87 & 11.25 & 803a  & 1969  & 2014  & 46 \\
    Ny-Alesund & Norway & Europe & 78.93 & 11.95 & 823a  & 1976  & 2014  & 39 \\
    Vardo & Norway & Europe & 70.33 & 31.10 & 805a  & 1979  & 2014  & 36 \\
    Balboa & Panama & South America & 9.10  & -79.63 & 302a  & 1907  & 2014  & 108 \\
    Cristobal & Panama & South America & 9.40  & -80.05 & 266a  & 1907  & 2014  & 108 \\
    Rabaul & Papua New Guinea & SIDS  & -4.20 & 152.25 & 010a  & 1966  & 1997  & 32 \\
    Lobos de Afuera & Peru  & South America & -6.93 & -80.72 & 084a  & 1982  & 2014  & 33 \\
    Callao-B & Peru  & South America & -12.08 & -77.17 & 093b  & 1970  & 2015  & 46 \\
    Legaspi & Philippines & Pacific & 13.25 & 123.83 & 371a  & 1984  & 2015  & 32 \\
    Manila & Philippines & Pacific & 14.60 & 120.98 & 370a  & 1984  & 2015  & 32 \\
    Cascais & Portugal & Europe & 38.77 & -9.42 & 209a  & 1959  & 2005  & 47 \\
    Funchal-B & Portugal & Europe & 32.73 & -17.03 & 218b  & 1982  & 2013  & 32 \\
    Kanton-B & Rep. of Kiribati & SIDS  & -2.90 & -171.73 & 013b  & 1972  & 2012  & 41 \\
    Christmas-B & Rep. of Kiribati & SIDS  & 2.12  & -157.52 & 011b  & 1974  & 2015  & 42 \\
    Majuro-A & Rep. of Marshall I & SIDS  & 7.17  & 171.43 & 005a  & 1968  & 1999  & 32 \\
    Kwajalein & Rep. of Marshall I & SIDS  & 8.73  & 167.73 & 055a  & 1946  & 2014  & 69 \\
    Malakal-B & Republic of Belau & SIDS  & 7.45  & 134.58 & 007b  & 1969  & 2015  & 47 \\
    Kaohsiung & Republic of China & Pacific & 22.75 & 120.33 & 340a  & 1980  & 2014  & 35 \\
    Keelung & Republic of China & Pacific & 25.22 & 121.85 & 341a  & 1980  & 2014  & 35 \\
    Luderitz & South Africa & Africa & -26.65 & 15.15 & 702a  & 1958  & 1995  & 38 \\
    Saldahna Bay & South Africa & Africa & -33.02 & 17.95 & 703a  & 1982  & 2011  & 30 \\
    Simon's Town & South Africa & Africa & -34.18 & 18.43 & 221a  & 1959  & 1999  & 41 \\
    Port Nolloth & South Africa & Africa & -29.25 & 16.87 & 701a  & 1958  & 1997  & 40 \\
    Port Elizabeth & South Africa & Africa & -34.05 & 25.75 & 184a  & 1978  & 2014  & 37 \\
    La Coruna & Spain & Europe & 43.37 & -8.40 & 830a  & 1943  & 2013  & 71 \\
    Ceuta & Spain & Europe & 35.90 & -5.32 & 207a  & 1944  & 2013  & 70 \\
    Vigo  & Spain & Europe & 42.23 & -8.73 & 208a  & 1943  & 1990  & 48 \\
    Stockholm & Sweden & Europe & 59.40 & 18.22 & 826a  & 1889  & 2014  & 126 \\
    Goteborg-Torsh. & Sweden & Europe & 57.70 & 11.85 & 819a  & 1967  & 2014  & 48 \\
    Zanzibar & Tanzania & Africa & -6.20 & 39.25 & 151a  & 1984  & 2015  & 32 \\
    Ko Lak & Thailand & South Asia/ Australia & 11.90 & 99.82 & 328a  & 1985  & 2015  & 31 \\
    Stornoway & United Kingdom & Europe & 58.28 & -6.43 & 295a  & 1976  & 2010  & 35 \\
    Lerwick & United Kingdom & Europe & 60.20 & -1.20 & 293a  & 1959  & 2010  & 52 \\
    Faraday & United Kingdom & Europe & -65.25 & -64.27 & 700a  & 1978  & 2013  & 36 \\
    Gibraltar-A & United Kingdom & Europe & 36.13 & -5.35 & 289a  & 1961  & 1992  & 32 \\
    Bermuda-B & United Kingdom & Europe & 32.43 & -64.73 & 259b  & 1985  & 2014  & 30 \\
    Newlyn, Cornwall & United Kingdom & Europe & 50.12 & -5.62 & 294a  & 1915  & 2010  & 96 \\
    Seward-C,  AK & USA   & Canada & 60.15 & -149.52 & 560c  & 1967  & 2014  & 48 \\
    Ketchikan,  AK & USA   & Canada & 55.33 & -131.70 & 571a  & 1937  & 2014  & 78 \\
    Valdez,  AK & USA   & Canada & 61.20 & -146.47 & 562a  & 1973  & 2014  & 42 \\
    Yakutat,  AK & USA   & Canada & 59.68 & -139.75 & 570a  & 1961  & 2014  & 54 \\
    Seldovia,  AK & USA   & Canada & 59.50 & -151.75 & 561a  & 1975  & 2014  & 40 \\
    Sitka,  AK & USA   & Canada & 57.07 & -135.42 & 559a  & 1938  & 2014  & 77 \\
    Sand Point,  AK & USA   & Canada & 55.37 & -160.52 & 574a  & 1973  & 2014  & 42 \\
    Dutch Harbor-B,  AK & USA   & Canada & 54.00 & -166.57 & 041b  & 1982  & 2014  & 33 \\
    Cordova-B,  AK & USA   & Canada & 60.63 & -145.78 & 583b  & 1964  & 2014  & 51 \\
    Kodiak Isl.,  AK & USA   & Canada & 57.87 & -152.62 & 039a  & 1975  & 2014  & 40 \\
    Adak,  AK & USA   & Canada & 51.98 & -176.77 & 040a  & 1950  & 2014  & 65 \\
    San Francisco,  CA & USA   & USA West & 37.87 & -122.60 & 551a  & 1897  & 2014  & 118 \\
    San Diego,  CA & USA   & USA West & 32.83 & -117.23 & 569a  & 1906  & 2014  & 109 \\
    Los Angeles,  CA & USA   & USA West & 33.75 & -118.32 & 567a  & 1923  & 2014  & 92 \\
    Crescent City,  CA & USA   & USA West & 41.85 & -124.18 & 556a  & 1933  & 2014  & 82 \\
    Monterey,  CA & USA   & USA West & 36.65 & -121.93 & 555a  & 1973  & 2014  & 42 \\
    Port San Luis,  CA & USA   & USA West & 35.27 & -120.85 & 565a  & 1948  & 2014  & 67 \\
    Santa Monica,  CA & USA   & USA West & 34.08 & -118.50 & 578a  & 1973  & 2014  & 42 \\
    La Jolla,  CA & USA   & USA West & 32.87 & -117.32 & 554a  & 1924  & 2014  & 91 \\
    New London,  CT & USA   & USA East & 41.40 & -72.12 & 744a  & 1938  & 2014  & 77 \\
    Lewes,  DE & USA   & USA East & 38.92 & -75.15 & 747a  & 1957  & 2014  & 58 \\
    Fernandina Beach,  FL & USA   & USA East & 30.72 & -81.47 & 240a  & 1897  & 1930  & 34 \\
    St. Petersburg,  FL & USA   & USA East & 27.85 & -82.72 & 759a  & 1946  & 2014  & 69 \\
    Pensacola,  FL & USA   & USA East & 30.43 & -87.33 & 762a  & 1923  & 2014  & 92 \\
    Mayport,  FL & USA   & USA East & 30.50 & -81.57 & 753a  & 1928  & 2000  & 73 \\
    Limetree Bay,  FL & USA   & USA East & 17.82 & -64.78 & 254a  & 1982  & 2014  & 33 \\
    Key West,  FL & USA   & USA East & 24.58 & -81.88 & 242a  & 1913  & 2014  & 102 \\
    Fort Pulaski,  GA & USA   & USA East & 32.03 & -80.92 & 752a  & 1935  & 2014  & 80 \\
    Hilo,  HI & USA   & Pacific & 19.73 & -155.07 & 060a  & 1927  & 2014  & 88 \\
    French Frigate,  HI & USA   & Pacific & 23.88 & -166.33 & 014a  & 1974  & 2007  & 34 \\
    Kahului,  HI & USA   & Pacific & 20.90 & -156.47 & 059a  & 1950  & 2014  & 65 \\
    Mokuoloe,  HI & USA   & Pacific & 21.43 & -157.80 & 061a  & 1957  & 2014  & 58 \\
    Honolulu-B,  HI & USA   & Pacific & 21.37 & -157.87 & 057b  & 1905  & 2014  & 110 \\
    Nawiliwili,  HI & USA   & Pacific & 21.97 & -159.35 & 058a  & 1954  & 2014  & 61 \\
    Grand Isle,  LA & USA   & USA East & 29.38 & -90.02 & 765a  & 1980  & 2014  & 35 \\
    Woods Hole,  MA & USA   & USA East & 41.58 & -70.72 & 742a  & 1957  & 2014  & 58 \\
    Nantucket,  MA & USA   & USA East & 41.30 & -70.22 & 743a  & 1965  & 2014  & 50 \\
    Boston,  MA & USA   & USA East & 42.40 & -71.07 & 741a  & 1921  & 2014  & 94 \\
    Portland,  ME & USA   & USA East & 43.72 & -70.37 & 252a  & 1910  & 2014  & 105 \\
    Eastport,  ME & USA   & USA East & 44.93 & -67.00 & 740a  & 1929  & 2014  & 86 \\
    Duck Pier,  NC & USA   & USA East & 36.18 & -75.87 & 260a  & 1978  & 2014  & 37 \\
    Wilmington,  NC & USA   & USA East & 34.32 & -77.98 & 750a  & 1935  & 2014  & 80 \\
    Atlantic City,  NJ & USA   & USA East & 39.40 & -74.43 & 264a  & 1911  & 2014  & 104 \\
    Cape May,  NJ & USA   & USA East & 38.98 & -75.05 & 746a  & 1965  & 2014  & 50 \\
    Montauk,  NY & USA   & USA East & 41.18 & -72.05 & 279a  & 1959  & 2014  & 56 \\
    New York,  NY & USA   & USA East & 40.70 & -74.15 & 745a  & 1920  & 2014  & 95 \\
    Charleston,  OR & USA   & USA West & 43.45 & -124.37 & 575a  & 1978  & 2014  & 37 \\
    South Beach,  OR & USA   & USA West & 44.70 & -124.13 & 592a  & 1967  & 2014  & 48 \\
    Astoria,  OR & USA   & USA West & 46.28 & -123.77 & 572a  & 1925  & 2014  & 90 \\
    Newport,  RI & USA   & USA East & 41.55 & -71.42 & 253a  & 1930  & 2014  & 85 \\
    Charleston,  SC & USA   & USA East & 32.92 & -80.00 & 261a  & 1921  & 2014  & 94 \\
    Rockport,  TX & USA   & USA East & 28.07 & -97.17 & 769a  & 1944  & 2014  & 71 \\
    Port Isabel,  TX & USA   & USA East & 26.15 & -97.35 & 772a  & 1977  & 2014  & 38 \\
    Chesapeake BBT,  VA & USA   & USA East & 36.97 & -76.23 & 749a  & 1975  & 2014  & 40 \\
    Neah Bay,  WA & USA   & USA East & 48.38 & -124.62 & 558a  & 1934  & 2014  & 81 \\
    Willapa Bay,  WA & USA   & USA East & 46.78 & -124.10 & 564a  & 1972  & 2014  & 43 \\
    Galveston, Pier 21, TX & USA   & USA East & 29.40 & -94.88 & 775a  & 1904  & 2014  & 111 \\
    Galveston, P. Pier, TX & USA   & USA East & 29.32 & -94.85 & 767a  & 1957  & 2011  & 55 \\
    Apra Harbor, Guam & USA Trust & Pacific & 13.43 & 144.65 & 053a  & 1948  & 2014  & 67 \\
    Wake  & USA Trust & Pacific & 19.28 & 166.62 & 051a  & 1950  & 2014  & 65 \\
    Johnston & USA Trust & Pacific & 16.78 & -169.65 & 052a  & 1947  & 2015  & 69 \\
    Midway & USA Trust & Pacific & 28.22 & -177.37 & 050a  & 1947  & 2014  & 68 \\
    Pago Pago & USA Trust & Pacific & -14.28 & -170.68 & 056a  & 1948  & 2014  & 67 \\
    Charlotte Amalie, VI & USA Trust & SIDS  & 18.35 & -64.95 & 255a  & 1978  & 2014  & 37 \\
    San Juan, PR & USA Trust & SIDS  & 18.55 & -66.12 & 245a  & 1977  & 2014  & 38 \\
    Magueyes Island, PR & USA Trust & SIDS  & 18.02 & -67.17 & 246a  & 1965  & 2014  & 50 \\
   \hline
\label{Stab:tgauges}
\end{longtable}}

\newpage

\begin{table}[htbp]
  \centering
\caption{Inventory of CMIP5 models and RCPs used for 1.5 $^\circ$C, 2.0 $^\circ$C, and 2.5 $^\circ$C global mean surface temperature (GMST) stabilization targets. Information is given for the 19-yr running average 2100 GMST ($^\circ$C; relative to 1875--1900) and the lengths of the GMST projections and the contributions to sea-level change from oceanographic processes and glacial ice (GIC;  from \citet{Marzeion2012}). `Local Ocean' is the local sea surface height above the geoid (i.e., 'zos') and `Thermal Expansion' refers to the contribution to the change in the global mean sea level due to thermal expansion (i.e.,'zostoga').}
    {\small
    \begin{tabular}{lrc|cccc}
    \multicolumn{7}{c}{1.5 $^\circ$C} \\
    Model & RCP & 2100 GMST ($^\circ$C)  & GMST  & Local Ocean & Thermal Expansion & GIC \\
    \hline
    bcc-csm1-1 & RCP 2.6 & 1.51  & 23    & 23    & 23    & 23 \\
    BNU-ESM & RCP 2.6 & 1.62  & 21    &       &       &  \\
    CCSM4 & RCP 2.6 & 1.45  & 23    & 23    & 21    & 21 \\
    FIO-ESM  & RCP 4.5 & 1.7   & 21    & 21    &       &  \\
    GFDL-ESM2G  & RCP 4.5 & 1.61  & 21    & 21    &       &  \\
    HadGEM2-AO & RCP 2.6 & 1.74  & 21    &       &       &  \\
    IPSL-CM5A-LR & RCP 2.6  & 1.74  & 23    & 23    & 23    & 23 \\
    IPSL-CM5A-MR & RCP 2.6 & 1.6   & 21    & 21    & 21    &  \\
    MIROC5 & RCP 2.6 & 1.62  & 21    &       &       & 21 \\
    MPI-ESM-LR & RCP 2.6 & 1.44  & 23    & 23    & 23    & 23 \\
    MPI-ESM-MR & RCP 2.6 & 1.36  & 21    & 21    & 21    &  \\
    MRI-CGCM3 & RCP 2.6 & 1.72  & 21    & 21    & 21    & 21 \\
    NorESM1-M & RCP 2.6 & 1.52  & 21    & 21    & 21    & 21 \\
    NorESM1-ME & RCP 2.6 & 1.67  & 21    & 21    & 21    &  \\
    \\
    \multicolumn{7}{c}{2.0 $^\circ$C} \\
   Model & RCP & 2100 GMST ($^\circ$C)  & GMST  & Local Ocean  & Thermal Expansion & GIC \\
    \hline
    bcc-csm1-1  & RCP 4.5 & 2.21  & 23    &    &       & 23 \\
    bcc-csm1-1-m  & RCP 4.5 & 2.13  & 21    & 21    & 21    &  \\
    CanESM2 & RCP 2.6  & 2.17  & 23    & 23    & 23    & 23 \\
    CESM1-BGC  & RCP 4.5 & 2.24  & 21    & 21    &       &  \\
    CESM1-CAM5  & RCP 2.6  & 2.13  & 23    &       &       &  \\
    CSIRO-MK3-6-0 & RCP 2.6  & 2.04  & 21    &       & 21    &  \\
    FGOALS-G2  & RCP 4.5 & 2.01  & 23    &       &       &  \\
    GFDL-ESM2M  & RCP 4.5 & 1.84  & 22    & 21    & 21    &  \\
    GISS-E2-H-CC & RCP 4.5 & 2.03  & 21    &       &       &  \\
    GISS-E2-R  & RCP 4.5 & 1.88  & 23    & 23    & 23    & 23 \\
    GISS-E2-R-CC  & RCP 4.5 & 1.87  & 21    & 21    & 21    &  \\
    HadGEM2-ES  & RCP 2.6  & 1.97  & 23    & 23    & 23    & 23 \\
    inmcm4  & RCP 4.5 & 2.04  & 21    & 21    & 21    & 21 \\
    \\
    \multicolumn{7}{c}{2.5 $^\circ$C} \\
      Model & RCP & 2100 GMST ($^\circ$C)  & GMST  & Local Ocean  & Thermal Expansion & GIC \\
     \hline
    CCSM4  & RCP 4.5 & 2.31  & 23    & 23    & 21    & 21 \\
    CNRM-CM5  & RCP 4.5 & 2.56  & 23    & 23    & 23    & 23 \\
    FIO-ESM & RCP 6.0 & 2.34  & 21    & 21    &       &  \\
    GFDL-CM3  & RCP 2.6 & 2.57  & 21    & 21    & 21    & 21 \\
    GFDL-ESM2G  & RCP 6.0 & 2.35  & 21    & 21    & 21    &  \\
    GFDL-ESM2M & RCP 6.0 & 2.53  & 21    & 21    & 21    &  \\
    GISS-E2-R & RCP 6.0 & 2.52  & 21    & 21    & 21    & 21 \\
    IPSL-CM5B-LR  & RCP 4.5 & 2.37  & 21    & 21    &       &  \\
    MIROC-ESM & RCP 2.6 & 2.32  & 21    & 21    & 21    & 21 \\
    MIROC-ESM-CHEM  & RCP 2.6 & 2.42  & 21    & 21    & 21    &  \\
    MIROC5  & RCP 4.5 & 2.38  & 21    &      &   21    & 21 \\
    MPI-ESM-LR  & RCP 4.5 & 2.38  & 23    & 23    & 23    & 23 \\
    MPI-ESM-MR  & RCP 4.5 & 2.39  & 21    & 21    & 21    &  \\
    MRI-CGCM3  & RCP 4.5 & 2.51  & 21    & 21    &       & 21 \\
    NorESM1-M & RCP 6.0 & 2.74  & 21    & 21    & 21    & 23 \\
    NorESM1-M  & RCP 4.5 & 2.33  & 23    & 23    & 21    &  \\
    NorESM1-ME  & RCP 4.5 & 2.44  & 21    & 21    & 21  & \\
    \hline
    \multicolumn{5}{l}{21 = to 2100, 22 = to 2200, 23 = to 2300}
    \end{tabular}}%
  \label{Stab:model_inventory}%
\end{table}%

\newpage 

\begin{table}[h]
\centering
\setlength{\tabcolsep}{4pt}
\caption{GMSL projections from a 1.75 $^\circ$C and a 2.25 $^\circ$C GMST scenario. All values are cm above 2000 CE baseline. AIS = Antarctic Ice Sheet, GIS = Greenland Ice Sheet; TE = Thermal Expansion; GIC = Glacial Ice Melt; LWS = Land-Water Storage. }
{\small
\begin{tabular}{l|ccc|ccc|}
&\multicolumn{3}{c|}{1.75$^\circ$C}&\multicolumn{3}{c|}{2.25$^\circ$C}  \\
cm&50   &  17--83   &  5--95  &50   &  17--83   &  5--95  \\
\hline  \multicolumn{7}{l}{2100---Components}  \\
AIS&6&-4--17&-8--35&6&-4--17&-8--35  \\
GIS&6&4--12&3--17&6&4--12&3--17  \\
TE&21&12--30&6--37&23&20--27&17--30  \\
GIC&12&8--15&6--17&13&9--16&6--20  \\
LWS&5&3--7&2--8&5&3--7&2--8  \\
\hline
Total&51&36--70&27--90&54&41--70&34--90  \\
\hline  \multicolumn{7}{l}{Projections  by  year}  \\
2050&25&20--30&17--34&25&21--30&18--34  \\
2070&35&27--45&22--54&37&30--45&26--54  \\
2100&51&36--70&27--90&54&41--70&34--90  \\
2150&73&44--110&30--158&80&51--115&38--163  \\
2200&98&46--164&21--252&107&55--172&30--260  \\
\end{tabular}}%
 \label{Stab:sensitivity}%
\end{table}%

\newpage

\newpage 


\begin{landscape}
\centering
\setlength{\tabcolsep}{1pt}
{\small
\begin{longtable}{llcccc|ccc|ccc}
\caption{Expected flood amplification factors (AF) for the 10-yr flood for 2050, 2100, and 2150 under 1.5$^\circ$C, 2.0$^\circ$C, and 2.5$^\circ$C global mean surface  temperature stabilization scenarios.} \\
& & \multicolumn{9}{c}{10-yr Flood} \\
& & & \multicolumn{3}{c}{2050} & \multicolumn{3}{c}{2100} & \multicolumn{3}{c}{2150} \\ \cmidrule(lr){4-6}  \cmidrule(lr){7-9} \cmidrule(lr){10-12} 
Site & Region & Historical Height (m above MHHW)&AF 1.5$^\circ$C & AF 2.0$^\circ$C & AF 2.5$^\circ$C&AF 1.5$^\circ$C & AF 2.0$^\circ$C & AF 2.5$^\circ$C&AF 1.5$^\circ$C & AF 2.0$^\circ$C & AF 2.5$^\circ$C \\
\hline
\endfirsthead
\multicolumn{9}{c}
{{\bfseries \tablename\ \thetable{} -- continued from previous page}} \\
& & \multicolumn{9}{c}{10-yr Flood} \\
& & & \multicolumn{3}{c}{2050} & \multicolumn{3}{c}{2100} & \multicolumn{3}{c}{2150} \\ \cmidrule(lr){4-6}  \cmidrule(lr){7-9} \cmidrule(lr){10-12} 
Site & Region & Historical Height (m above MHHW)&AF 1.5$^\circ$C & AF 2.0$^\circ$C & AF 2.5$^\circ$C&AF 1.5$^\circ$C & AF 2.0$^\circ$C & AF 2.5$^\circ$C&AF 1.5$^\circ$C & AF 2.0$^\circ$C & AF 2.5$^\circ$C \\
\hline
\endhead
\hline
\multicolumn{9}{c}{{Continued on next page}} \\
\endfoot
\endlastfoot
Buenos Aires&Argentina&2.15&2.1&2.3&2.3&8.1&12.6&13.8&57.3&100.5&95.5 \\
Fort Denison&Australia&0.63&27.7&57.0&48.6&460.2&746.8&872.3&921.2&1286.3&1314.1 \\
Bundaberg&Australia&0.91&15.7&23.8&21.5&232.9&397.4&460.0&717.8&1044.0&1030.3 \\
Brisbane&Australia&0.65&44.6&72.3&63.6&570.1&870.8&948.4&1035.1&1351.8&1335.2 \\
Spring Bay&Australia&0.57&46.2&83.1&98.6&704.6&1054.7&1143.0&1184.4&1464.5&1495.8 \\
Townsville&Australia&1.18&10.6&14.0&13.2&112.1&176.1&221.0&470.9&748.1&734.8 \\
Broome&Australia&2.27&11.1&12.0&13.2&39.2&49.7&56.3&125.7&187.1&176.6 \\
Cocos&Australia&0.51&180.9&238.1&330.0&1304.3&1519.5&1575.3&1596.9&1658.1&1701.9 \\
Darwin&Australia&1.53&18.5&20.0&21.9&88.7&121.7&142.1&322.4&492.4&478.1 \\
Esperance&Australia&0.74&23.0&29.9&39.1&453.8&647.9&787.6&1012.8&1276.6&1312.8 \\
Fremantle&Australia&0.74&22.5&30.4&39.3&469.0&696.8&842.2&1094.9&1335.6&1397.0 \\
Cananeia&Brazil&0.96&19.3&28.6&26.8&418.1&681.5&755.7&1124.4&1311.9&1381.4 \\
Ilha Fiscal, RJ&Brazil&0.83&15.1&22.9&21.0&335.6&585.5&652.1&921.1&1162.0&1208.7 \\
Victoria, BC&Canada&0.91&4.7&5.9&7.0&101.1&131.5&183.4&370.7&576.4&542.8 \\
Prince Rupert&Canada&1.56&4.8&5.2&5.4&38.7&43.8&50.7&154.7&165.1&208.8 \\
Tofino&Canada&1.11&1.7&2.1&2.4&30.5&37.9&50.5&128.7&197.3&180.0 \\
St. John's-A&Canada&0.82&23.6&38.1&42.3&278.8&639.0&639.7&660.0&1075.5&1042.2 \\
Halifax&Canada&0.84&34.5&48.1&57.2&437.5&842.7&911.5&965.0&1275.9&1392.6 \\
Churchill&Canada&1.28&0.1&0.2&0.1&1.3&4.4&3.3&10.9&11.0&10.4 \\
Puerto Montt&Chile&1.60&7.0&8.1&9.3&32.7&44.0&52.2&120.4&199.1&192.4 \\
Juan Fernandez-B&Chile&0.52&32.3&44.5&53.1&579.7&759.3&911.6&1024.5&1239.3&1267.6 \\
Antofagasta&Chile&0.48&20.1&38.1&43.1&453.7&684.4&807.5&862.0&1126.6&1157.5 \\
Easter-C&Chile&0.59&11.2&18.4&18.8&510.5&699.6&822.7&1031.0&1249.9&1270.3 \\
Valparaiso&Chile&0.53&13.4&22.2&24.4&262.2&452.5&551.9&658.9&939.8&964.9 \\
Xiamen&China&1.43&6.3&7.3&8.7&71.2&91.9&129.0&313.5&489.1&502.5 \\
Buenaventura&Colombia&1.06&22.2&27.1&27.5&194.8&293.9&342.2&638.9&910.9&894.4 \\
Tumaco&Colombia&0.88&10.7&15.5&15.8&121.8&195.1&233.8&402.2&649.7&623.8 \\
Cartagena&Colombia&0.25&1816.1&1747.8&1799.9&1825.7&1818.2&1826.0&1822.6&1808.5&1823.4 \\
Penrhyn&Cook Islands&0.34&437.7&609.7&621.2&1549.6&1630.4&1672.8&1632.5&1688.9&1718.3 \\
Quepos-A&Costa Rica&0.75&32.4&43.6&46.1&481.4&686.4&793.5&1019.8&1263.8&1274.2 \\
Hornbaek&Denmark&1.25&4.3&42.8&3.8&37.5&287.1&45.7&171.1&347.4&190.8 \\
Gedser&Denmark&1.29&3.8&23.4&3.4&40.4&228.1&51.0&201.0&315.9&237.0 \\
Baltra-B&Ecuador&0.71&18.9&29.3&31.3&473.7&662.4&778.3&983.0&1224.1&1241.1 \\
Santa Cruz&Ecuador&0.63&35.0&54.5&57.5&656.1&856.2&985.3&1119.2&1333.8&1353.8 \\
La Libertad&Ecuador&0.74&51.1&73.6&75.6&954.1&1140.8&1276.9&1485.6&1587.5&1628.2 \\
Acajutla-A&El Salvador&0.63&55.8&78.7&86.4&827.4&1037.5&1157.1&1305.1&1465.5&1489.2 \\
Chuuk&Fd. St. Micronesia&0.40&202.5&274.0&298.8&1182.5&1390.0&1504.0&1417.4&1557.8&1596.3 \\
Kapingamarangi&Fd. St. Micronesia&0.50&122.1&155.9&191.6&1176.0&1317.1&1451.6&1449.1&1582.2&1602.5 \\
Pohnpei-B&Fd. St. Micronesia&0.51&127.3&166.3&175.5&1105.4&1310.7&1426.8&1418.6&1556.3&1587.8 \\
Yap-B&Fd. St. Micronesia&0.54&70.9&96.5&109.9&951.8&1270.6&1369.6&1370.6&1527.1&1557.8 \\
Suva-C&Fiji&0.51&269.0&317.3&333.3&1557.0&1513.7&1703.6&1731.1&1748.0&1720.1 \\
Noumea&France&0.43&208.1&307.2&291.3&1242.8&1415.9&1550.6&1452.0&1611.9&1572.7 \\
Brest&France&1.70&8.6&7.9&9.3&37.5&50.9&54.4&137.5&207.4&165.2 \\
Marseille&France&0.68&10.2&125.7&15.1&262.5&570.1&411.9&812.3&754.0&1046.2 \\
Rikitea&French Polynesia&0.31&532.3&751.2&761.0&1571.6&1664.8&1686.9&1625.9&1705.8&1673.2 \\
Papeete-B&French Polynesia&0.31&637.9&885.9&878.7&1672.5&1704.1&1762.1&1709.3&1752.2&1733.2 \\
Cuxhaven&Germany&2.65&2.1&2.1&2.0&5.7&12.4&6.3&36.4&34.7&34.0 \\
Malin Head&Ireland&1.27&4.6&5.0&3.8&32.5&56.9&34.8&135.6&128.7&138.0 \\
Hakodate&Japan&0.48&34.8&56.6&111.6&509.7&717.1&872.1&874.6&1123.6&1141.5 \\
Hamada&Japan&0.63&27.3&42.3&80.3&668.0&896.5&1030.5&1114.0&1359.0&1331.2 \\
Maisaka&Japan&0.77&1.8&2.2&3.8&74.2&109.8&179.4&284.6&478.8&522.3 \\
Ishigaki&Japan&0.78&22.7&31.9&54.8&627.5&784.4&958.3&1126.2&1336.7&1359.5 \\
Naha&Japan&0.71&40.0&57.7&105.2&748.2&912.5&1111.4&1192.0&1382.8&1416.4 \\
Toyama&Japan&0.47&172.8&260.0&421.9&1308.0&1476.0&1520.1&1521.7&1649.8&1609.0 \\
Hosojima&Japan&0.81&6.1&8.2&13.8&196.5&305.1&415.5&575.4&846.4&872.8 \\
Kushiro&Japan&0.52&1570.4&1537.4&1584.6&1826.1&1823.6&1826.1&1826.1&1822.4&1826.0 \\
Abashiri&Japan&0.64&38.7&80.8&93.6&664.3&905.2&1014.8&1164.8&1334.2&1363.9 \\
Mera&Japan&0.60&155.1&207.1&311.1&1511.8&1613.3&1615.1&1724.9&1747.7&1710.9 \\
Wakkanai&Japan&0.53&125.1&215.6&312.1&1265.2&1491.4&1488.6&1563.2&1678.3&1639.2 \\
Chichijima&Japan&0.57&107.0&229.6&321.6&1199.0&1303.5&1488.0&1522.3&1600.6&1621.0 \\
Nishinoomote&Japan&0.71&38.5&51.8&92.1&710.7&902.5&1044.5&1182.9&1408.9&1397.1 \\
Naze&Japan&0.73&63.0&79.5&139.1&947.6&1092.5&1288.0&1361.0&1520.9&1540.4 \\
Hachinohe&Japan&0.50&396.3&509.4&671.4&1651.2&1743.0&1729.8&1779.9&1783.0&1770.1 \\
Miyakejima&Japan&0.87&157.1&173.4&242.2&1658.2&1711.9&1701.7&1813.4&1806.8&1798.8 \\
Nakano Shima&Japan&0.74&51.0&65.7&114.1&858.9&1017.1&1156.1&1297.8&1484.3&1464.0 \\
Ofunato&Japan&0.50&1221.5&1376.5&1306.1&1824.9&1823.2&1819.2&1825.7&1823.4&1822.1 \\
Nagasaki&Japan&0.83&29.8&37.0&49.8&403.9&565.3&687.5&873.0&1144.7&1140.9 \\
Aburatsu&Japan&0.82&10.3&13.9&25.9&350.3&509.0&646.2&859.3&1137.3&1148.4 \\
Kushimoto&Japan&0.68&85.6&112.7&182.0&1254.5&1435.6&1473.5&1610.4&1686.8&1667.5 \\
Cendering&Malaysia&0.90&13.7&16.9&19.4&232.5&351.1&445.7&654.3&889.4&886.5 \\
Johor Baharu&Malaysia&0.81&34.1&40.8&46.9&432.4&589.6&717.2&898.4&1130.0&1133.7 \\
Kuantan&Malaysia&0.96&18.4&21.8&24.6&253.0&370.3&468.0&706.0&940.5&938.7 \\
Keling&Malaysia&0.65&40.1&49.9&58.5&518.0&679.3&817.1&917.1&1136.6&1139.2 \\
Lumut&Malaysia&0.76&25.8&33.4&40.7&407.7&563.3&672.2&832.6&1068.3&1060.1 \\
Kelang&Malaysia&1.22&12.5&13.9&15.7&114.1&156.9&204.5&425.4&599.5&588.5 \\
Langkawi&Malaysia&0.83&29.0&35.9&42.9&322.8&461.0&562.0&736.7&973.8&964.1 \\
Penang&Malaysia&0.72&46.8&59.6&72.8&535.6&713.5&827.1&954.6&1183.7&1178.7 \\
Port Louis-C&Mauritius&0.41&109.6&177.3&271.5&1038.2&1329.3&1377.1&1320.4&1522.8&1542.5 \\
Rodrigues&Mauritius&0.69&27.8&41.7&56.5&699.4&919.4&1050.9&1229.9&1454.6&1461.8 \\
Manzanillo-A&Mexico&0.52&261.1&377.0&405.5&1670.8&1692.5&1737.9&1777.3&1760.1&1776.8 \\
Ensenada&Mexico&0.65&53.6&78.9&84.2&694.5&921.0&1028.5&1226.9&1387.9&1419.1 \\
Salina Cruz&Mexico&0.53&126.7&183.0&195.9&1313.4&1456.9&1552.7&1581.8&1655.7&1677.6 \\
Acapulco-A, Gro.&Mexico&0.51&437.5&596.6&645.7&1800.8&1782.9&1807.9&1819.0&1799.7&1811.1 \\
Cabo San Lucas&Mexico&0.59&49.2&73.9&76.7&724.4&947.7&1051.6&1195.4&1376.2&1395.0 \\
Guaymas&Mexico&0.52&180.4&255.8&346.6&1577.0&1639.3&1663.3&1746.1&1742.9&1761.3 \\
Saipan-B&N. Mariana Islands&0.44&125.2&186.6&209.5&1187.6&1311.4&1479.6&1398.0&1511.3&1584.4 \\
Marsden Point&New Zealand&0.75&8.2&28.3&21.9&335.2&649.1&714.4&866.0&1254.1&1240.3 \\
Tauranga&New Zealand&0.51&62.0&199.0&155.5&949.0&1307.1&1401.9&1272.3&1588.4&1596.8 \\
Taranaki&New Zealand&1.10&3.8&7.8&6.5&88.7&194.1&214.7&439.4&757.9&733.1 \\
Wellington&New Zealand&0.49&140.0&371.0&292.0&1311.5&1564.6&1626.5&1519.4&1720.1&1731.8 \\
Tregde&Norway&0.77&11.5&17.5&9.8&112.7&215.3&143.5&350.2&407.7&450.3 \\
Rorvik&Norway&1.15&1.5&3.2&1.7&17.2&36.0&22.3&77.5&86.1&76.7 \\
Ny-Alesund&Norway&0.66&0.6&18.3&0.0&6.4&21.7&5.5&51.8&23.3&32.2 \\
Vardo&Norway&1.03&3.7&15.1&5.7&37.5&111.6&64.6&135.6&234.3&184.0 \\
Balboa&Panama&1.24&17.1&18.9&20.2&106.9&169.1&189.2&413.3&668.1&654.1 \\
Cristobal&Panama&0.29&915.3&937.3&1108.2&1642.4&1711.5&1787.3&1635.2&1718.8&1750.3 \\
Rabaul&Papua New Guinea&0.36&293.7&371.7&446.8&1360.1&1467.7&1597.7&1511.0&1628.0&1649.2 \\
Lobos de Afuera&Peru&0.60&22.3&41.6&42.5&614.8&815.8&937.3&1060.7&1282.5&1300.7 \\
Callao-B&Peru&0.46&29.3&57.1&66.4&687.4&885.0&1029.1&1054.0&1281.4&1307.1 \\
Legaspi&Philippines&0.51&288.4&416.7&464.0&1616.7&1669.6&1742.3&1746.6&1740.0&1778.1 \\
Manila&Philippines&0.69&424.1&546.3&613.1&1821.7&1815.2&1824.0&1825.5&1819.2&1825.5 \\
Cascais&Portugal&0.93&17.1&15.4&21.6&156.3&236.4&313.1&556.8&790.4&802.6 \\
Funchal-B&Portugal&0.64&93.9&84.8&129.6&739.2&930.6&1056.5&1170.9&1354.8&1459.1 \\
Kanton-B&Rep. of Kiribati&0.43&122.4&191.0&199.9&1178.8&1349.5&1406.3&1409.2&1538.2&1577.7 \\
Christmas-B&Rep. of Kiribati&0.42&209.0&322.3&333.7&1405.8&1490.2&1540.0&1541.7&1629.3&1659.5 \\
Majuro-A&Rep. of Marshall I&0.60&110.2&139.8&137.8&1019.3&1207.0&1332.9&1396.9&1538.6&1577.0 \\
Kwajalein&Rep. of Marshall I&0.51&222.8&274.2&283.6&1275.1&1460.0&1544.3&1533.4&1617.6&1669.9 \\
Malakal-B&Republic of Belau&0.51&139.2&177.0&199.6&1069.9&1381.6&1450.4&1446.9&1590.3&1615.5 \\
Kaohsiung&Republic of China&0.58&25.5&38.2&54.8&482.7&634.2&778.0&818.2&1049.7&1062.1 \\
Keelung&Republic of China&0.67&24.0&36.9&68.0&774.9&945.7&1081.2&1174.9&1377.8&1376.7 \\
Luderitz&South Africa&0.55&87.6&108.3&131.5&996.0&1228.5&1371.7&1350.3&1521.8&1581.8 \\
Saldahna Bay&South Africa&0.60&79.4&93.1&113.5&825.8&1066.8&1184.7&1233.7&1438.8&1484.8 \\
Simon's Town&South Africa&0.62&75.5&90.1&111.3&934.4&1185.8&1309.2&1371.6&1542.0&1593.1 \\
Port Nolloth&South Africa&0.64&57.8&69.8&75.9&750.8&1000.9&1142.8&1222.7&1441.6&1499.4 \\
Port Elizabeth&South Africa&0.76&24.8&34.2&36.2&418.9&623.9&752.6&919.5&1211.9&1241.4 \\
La Coruna&Spain&1.07&23.4&21.1&25.3&157.3&222.3&263.8&559.7&751.3&757.2 \\
Ceuta&Spain&0.45&49.7&73.1&104.3&633.5&791.4&1005.8&1050.0&1264.2&1344.7 \\
Vigo&Spain&1.10&17.6&16.2&20.4&144.5&209.2&254.6&543.4&750.8&748.4 \\
Stockholm&Sweden&0.76&2.9&233.3&1.5&31.6&465.6&34.0&85.0&13.6&101.1 \\
Goteborg-Torsh.&Sweden&1.12&2.9&14.6&2.5&28.8&162.2&33.2&124.8&188.2&129.7 \\
Zanzibar&Tanzania&1.06&30.4&35.2&37.4&201.8&288.5&365.2&577.0&839.3&851.7 \\
Ko Lak&Thailand&0.93&30.1&36.0&40.0&413.9&576.0&690.9&950.9&1178.7&1181.8 \\
Stornoway&United Kingdom&1.39&7.0&6.5&6.1&35.7&49.6&38.3&141.8&123.8&139.4 \\
Lerwick&United Kingdom&0.84&9.8&12.3&8.1&90.6&136.9&100.6&289.2&288.7&326.8 \\
Faraday&United Kingdom&0.80&4.7&7.0&5.2&41.0&49.1&83.2&280.7&537.2&563.1 \\
Gibraltar-A&United Kingdom&0.43&49.2&69.5&99.3&508.8&671.9&872.4&899.4&1135.4&1213.4 \\
Bermuda-B&United Kingdom&0.53&72.5&91.2&121.9&927.8&1072.9&1152.6&1304.8&1454.9&1430.6 \\
Newlyn, Cornwall&United Kingdom&1.24&15.8&15.9&15.4&87.2&129.1&111.0&304.6&375.9&408.1 \\
Seward-C, AK&USA&1.28&0.5&0.8&0.6&9.7&12.6&11.2&51.0&58.3&59.5 \\
Ketchikan, AK&USA&1.53&2.5&2.6&2.8&22.6&25.7&29.6&93.7&97.8&123.3 \\
Valdez, AK&USA&1.29&0.2&0.4&0.3&6.2&8.6&6.9&35.1&42.6&40.5 \\
Yakutat, AK&USA&1.26&0.0&0.0&0.0&2.7&3.6&3.0&16.4&16.1&16.6 \\
Seldovia, AK&USA&1.79&0.1&0.1&0.1&0.7&0.8&0.7&10.7&11.3&10.9 \\
Sitka, AK&USA&1.21&1.0&1.0&1.2&17.3&19.9&22.4&76.7&76.6&98.1 \\
Sand Point, AK&USA&1.12&5.1&5.4&6.1&80.5&93.1&128.6&337.4&372.4&501.4 \\
Dutch Harbor-B, AK&USA&0.72&0.8&1.0&1.3&39.5&48.7&64.9&133.0&148.3&204.6 \\
Cordova-B, AK&USA&1.35&7.7&10.1&9.4&89.7&115.9&118.9&418.7&437.7&526.8 \\
Kodiak Isl., AK&USA&1.09&0.1&0.1&0.2&3.9&4.4&4.3&20.1&19.9&20.9 \\
Adak, AK&USA&0.78&4.7&6.0&6.6&91.0&119.3&160.8&278.8&310.4&421.3 \\
San Francisco, CA&USA&0.68&20.8&33.5&37.4&527.3&731.2&864.1&1099.2&1296.5&1323.2 \\
San Diego, CA&USA&0.68&56.7&80.5&86.6&681.7&910.2&1015.4&1245.3&1402.2&1436.1 \\
Los Angeles, CA&USA&0.66&32.1&49.2&52.4&412.0&605.2&712.9&906.8&1136.6&1159.9 \\
Crescent City, CA&USA&0.92&2.9&3.8&4.5&64.3&81.9&124.0&250.7&407.1&387.6 \\
Monterey, CA&USA&0.68&21.8&34.2&39.4&442.6&634.4&756.6&972.2&1193.7&1214.1 \\
Port San Luis, CA&USA&0.69&16.8&27.2&31.1&320.0&478.4&595.3&795.7&1039.5&1054.6 \\
Santa Monica, CA&USA&0.70&28.4&43.7&46.8&427.0&619.9&728.6&955.8&1176.1&1199.6 \\
La Jolla, CA&USA&0.66&56.2&82.2&88.7&713.8&945.6&1049.0&1264.4&1416.3&1449.4 \\
New London, CT&USA&1.04&5.6&8.0&8.3&141.0&329.5&336.4&540.5&934.0&953.3 \\
Fernandina Beach, FL&USA&0.92&13.1&15.0&18.6&238.7&458.4&506.3&799.6&1090.6&1157.7 \\
St. Petersburg, FL&USA&0.79&9.4&22.1&14.9&435.4&738.5&840.6&1082.9&1320.2&1384.4 \\
Pensacola, FL&USA&0.84&4.1&9.8&5.6&221.2&470.6&500.0&799.1&1105.9&1131.4 \\
Mayport, FL&USA&0.61&71.8&78.1&109.1&807.4&1136.5&1256.4&1293.1&1463.7&1587.6 \\
Limetree Bay, FL&USA&0.30&696.5&650.0&827.5&1634.2&1625.9&1730.4&1636.2&1674.1&1732.1 \\
Key West, FL&USA&0.43&276.4&324.1&439.4&1470.3&1517.0&1689.0&1609.6&1652.7&1713.1 \\
Fort Pulaski, GA&USA&0.76&53.5&56.3&77.1&666.0&1000.1&1109.0&1279.4&1462.2&1584.0 \\
Hilo, HI&USA&0.42&503.3&596.0&721.0&1686.5&1719.0&1743.5&1734.5&1760.8&1750.6 \\
French Frigate, HI&USA&0.38&362.6&527.7&579.9&1465.2&1610.7&1656.8&1550.5&1659.4&1656.6 \\
Kahului, HI&USA&0.36&650.1&742.3&873.4&1678.4&1710.4&1740.8&1706.5&1745.7&1736.7 \\
Mokuoloe, HI&USA&0.36&449.0&540.9&657.4&1566.5&1631.2&1675.3&1608.1&1697.0&1687.3 \\
Honolulu-B, HI&USA&0.36&442.6&534.6&650.3&1564.9&1630.0&1674.5&1607.2&1696.3&1686.8 \\
Nawiliwili, HI&USA&0.37&361.3&455.3&553.0&1486.1&1592.8&1628.3&1554.4&1665.9&1658.2 \\
Grand Isle, LA&USA&1.02&19.2&38.8&31.0&1604.4&1640.5&1762.9&1823.4&1800.8&1823.3 \\
Woods Hole, MA&USA&0.92&12.4&18.1&20.0&262.6&566.0&601.9&779.2&1118.9&1215.9 \\
Nantucket, MA&USA&0.95&8.3&13.5&14.0&264.8&559.9&605.6&834.6&1154.7&1270.9 \\
Boston, MA&USA&1.04&15.0&18.9&22.1&190.0&391.9&414.3&583.5&954.2&970.9 \\
Portland, ME&USA&0.93&19.2&25.9&29.1&196.8&418.7&432.0&530.0&915.5&902.5 \\
Eastport, ME&USA&1.27&18.8&23.3&26.5&112.2&236.1&230.3&323.3&667.9&535.3 \\
Duck Pier, NC&USA&0.90&23.8&34.7&43.8&461.1&844.5&908.1&1157.7&1380.0&1518.5 \\
Wilmington, NC&USA&0.58&47.3&95.6&104.2&798.3&1177.8&1268.7&1288.1&1468.2&1609.8 \\
Atlantic City, NJ&USA&0.94&24.0&34.8&43.2&448.6&833.0&909.3&1135.4&1372.4&1540.7 \\
Cape May, NJ&USA&0.88&25.9&33.8&43.4&465.9&882.1&954.3&1153.3&1379.6&1573.3 \\
Montauk, NY&USA&0.94&12.4&18.2&20.5&271.2&563.8&619.7&833.5&1161.1&1285.0 \\
New York, NY&USA&1.09&6.1&8.7&9.6&154.8&358.7&365.1&582.7&974.2&1010.1 \\
Charleston, OR&USA&0.96&6.7&8.2&9.7&124.2&160.1&238.2&462.9&675.5&658.9 \\
South Beach, OR&USA&1.04&8.9&10.7&12.2&154.1&196.8&281.1&575.5&805.1&787.0 \\
Astoria, OR&USA&1.00&3.3&4.2&4.8&58.8&73.7&106.6&230.1&377.0&346.3 \\
Newport, RI&USA&0.85&23.5&32.4&38.4&360.7&700.7&780.7&907.9&1206.2&1340.0 \\
Charleston, SC&USA&0.77&35.8&43.0&55.1&579.9&918.4&1023.3&1211.1&1430.0&1539.0 \\
Galveston (Pier 21), TX&USA&0.98&14.2&29.9&20.8&928.2&1222.7&1371.5&1679.3&1686.9&1761.8 \\
Rockport, TX&USA&0.63&141.8&236.8&247.7&1662.0&1699.0&1771.5&1797.2&1774.0&1813.2 \\
Port Isabel, TX&USA&0.64&40.6&78.5&73.8&1059.9&1320.0&1465.2&1517.5&1612.8&1695.4 \\
Galveston (P. Pier), TX&USA&1.14&8.2&15.5&10.9&546.7&872.5&984.8&1513.4&1590.8&1682.0 \\
Chesapeake BBT, VA&USA&1.05&12.0&16.8&19.5&289.1&600.5&652.1&998.4&1269.7&1413.2 \\
Neah Bay, WA&USA&1.11&1.3&1.7&2.0&24.7&30.9&41.8&105.5&160.2&142.6 \\
Willapa Bay, WA&USA&1.38&2.6&3.0&3.3&33.5&40.8&56.0&162.0&254.0&232.4 \\
Lewes, DE&USA&1.04&10.5&13.6&16.5&227.5&536.2&549.7&824.6&1158.6&1281.4 \\
Apra Harbor, Guam&USA Trust&0.29&852.2&920.5&1119.1&1620.4&1653.0&1759.7&1685.9&1700.2&1748.4 \\
Wake&USA Trust&0.52&107.9&206.3&179.3&1182.8&1349.6&1475.0&1432.9&1553.2&1594.8 \\
Johnston&USA Trust&0.54&58.7&98.8&91.2&957.9&1177.3&1250.8&1272.2&1458.8&1483.4 \\
Midway&USA Trust&0.61&24.2&58.8&67.2&758.4&966.6&1180.4&1185.1&1369.9&1407.6 \\
Pago Pago&USA Trust&0.38&579.2&722.6&777.5&1715.7&1760.7&1768.3&1752.0&1774.3&1771.4 \\
Charlotte Amalie, VI&USA Trust&0.31&618.5&584.8&740.8&1595.4&1590.4&1712.8&1619.7&1662.3&1726.1 \\
San Juan, PR&USA Trust&0.33&575.6&545.9&701.2&1586.4&1585.0&1713.7&1628.6&1670.1&1731.8 \\
Magueyes Island, PR&USA Trust&0.27&944.5&842.1&1046.0&1678.5&1642.5&1754.9&1659.1&1686.0&1740.4 \\
\hline
\end{longtable}}
\label{Stab:af_table_rp10}
\end{landscape}

\newpage

%
%

\begin{landscape}
\centering
\setlength{\tabcolsep}{1pt}
{\small
\begin{longtable}{llcccc|ccc|ccc}
\caption{Expected flood amplification factors (AF) for the 100-yr flood for 2050, 2100, and 2150 under 1.5$^\circ$C, 2.0$^\circ$C, and 2.5$^\circ$C global mean surface  temperature stabilization scenarios.} \\
& & \multicolumn{9}{c}{100-yr Flood} \\
& & & \multicolumn{3}{c}{2050} & \multicolumn{3}{c}{2100} & \multicolumn{3}{c}{2150} \\ \cmidrule(lr){4-6}  \cmidrule(lr){7-9} \cmidrule(lr){10-12} 
Site & Region & Historical Height (m above MHHW)&AF 1.5$^\circ$C & AF 2.0$^\circ$C & AF 2.5$^\circ$C&AF 1.5$^\circ$C & AF 2.0$^\circ$C & AF 2.5$^\circ$C&AF 1.5$^\circ$C & AF 2.0$^\circ$C & AF 2.5$^\circ$C \\
\hline
\endfirsthead
\multicolumn{9}{c}
{{\bfseries \tablename\ \thetable{} -- continued from previous page}} \\
& & \multicolumn{9}{c}{100-yr Flood} \\
& & & \multicolumn{3}{c}{2050} & \multicolumn{3}{c}{2100} & \multicolumn{3}{c}{2150} \\ \cmidrule(lr){4-6}  \cmidrule(lr){7-9} \cmidrule(lr){10-12} 
Site & Region & Historical Height (m above MHHW)&AF 1.5$^\circ$C & AF 2.0$^\circ$C & AF 2.5$^\circ$C&AF 1.5$^\circ$C & AF 2.0$^\circ$C & AF 2.5$^\circ$C&AF 1.5$^\circ$C & AF 2.0$^\circ$C & AF 2.5$^\circ$C \\
\hline
\endhead
\hline
\multicolumn{9}{c}{{Continued on next page}} \\
\endfoot
\endlastfoot
Buenos Aires&Argentina&3.07&1.6&1.7&1.7&3.8&5.2&5.6&95.7&149.6&128.9 \\
Fort Denison&Australia&0.74&74.0&189.0&152.8&2542.8&4577.0&5531.0&7203.0&10865.5&11223.7 \\
Bundaberg&Australia&1.24&6.1&10.8&9.1&508.6&887.6&1071.2&3179.6&5699.0&5496.6 \\
Brisbane&Australia&0.78&101.7&201.5&167.3&3001.5&5219.8&5830.1&7968.1&11411.9&11242.4 \\
Spring Bay&Australia&0.71&81.2&183.5&229.1&3382.9&6517.2&7076.8&9030.1&12442.5&12587.4 \\
Townsville&Australia&1.36&25.5&39.1&35.1&539.1&845.8&1074.4&3008.8&5213.4&5065.9 \\
Broome&Australia&2.41&40.1&47.1&53.5&266.1&351.1&401.4&972.6&1457.1&1368.6 \\
Cocos&Australia&0.58&849.8&1126.4&1689.8&10928.1&13440.3&14313.3&14987.4&15998.4&16422.2 \\
Darwin&Australia&1.69&43.1&55.8&64.5&531.6&740.8&868.4&2271.5&3566.9&3432.9 \\
Esperance&Australia&0.81&121.8&158.1&211.5&3223.6&4821.1&6073.0&8802.3&11631.6&11861.0 \\
Fremantle&Australia&0.89&46.3&67.2&93.1&2188.3&3628.9&4606.8&7868.5&10880.9&11144.6 \\
Cananeia&Brazil&1.37&6.4&9.2&8.4&558.2&1228.1&1272.9&4708.5&7576.4&7335.4 \\
Ilha Fiscal, RJ&Brazil&1.06&15.9&26.0&22.9&1025.3&2188.5&2342.6&5582.3&8439.0&8389.6 \\
Victoria, BC&Canada&1.06&10.4&14.1&17.8&540.0&687.0&963.8&2486.3&4057.3&3735.3 \\
Prince Rupert&Canada&1.74&10.6&11.9&12.6&210.8&242.4&276.1&1048.4&1107.4&1382.3 \\
Tofino&Canada&1.29&2.6&3.6&4.6&174.9&206.8&268.7&884.3&1293.4&1162.9 \\
St. John's-A&Canada&1.02&35.7&66.3&70.6&1166.0&3125.4&2932.8&3837.6&8127.2&6605.6 \\
Halifax&Canada&1.16&19.9&31.4&34.0&1096.0&2993.2&2634.5&4401.8&8623.0&7766.9 \\
Churchill&Canada&1.63&0.2&0.2&0.2&1.8&25.7&5.8&77.0&77.9&72.1 \\
Puerto Montt&Chile&1.70&22.0&28.1&34.3&209.1&300.2&364.5&936.8&1584.8&1519.9 \\
Juan Fernandez-B&Chile&0.61&89.6&140.9&172.3&3578.4&5278.6&6614.0&8541.8&11023.2&11217.8 \\
Antofagasta&Chile&0.60&28.3&71.9&83.8&2205.9&3906.6&4764.0&6434.8&9279.2&9479.0 \\
Easter-C&Chile&0.92&4.0&5.3&5.3&871.0&1390.8&1825.5&5196.1&7442.6&7908.2 \\
Valparaiso&Chile&0.61&36.2&71.8&80.2&1595.7&3000.7&3743.6&5312.1&8140.9&8277.1 \\
Xiamen&China&1.72&8.7&10.7&13.5&253.4&317.9&460.9&1588.4&2624.9&2734.7 \\
Buenaventura&Colombia&1.20&57.2&88.4&88.9&1093.6&1698.8&1978.8&4615.7&7169.2&6882.2 \\
Tumaco&Colombia&1.01&19.3&37.1&37.7&677.0&1112.8&1328.9&2873.0&4975.3&4675.1 \\
Cartagena&Colombia&0.30&17415.6&16094.6&17176.9&18238.4&18116.9&18253.9&18198.6&18021.3&18218.0 \\
Penrhyn&Cook Islands&0.53&244.9&464.1&384.5&9644.5&11494.0&12640.6&13623.0&15143.5&15438.9 \\
Quepos-A&Costa Rica&0.79&229.7&311.7&328.5&4000.0&5913.7&6889.0&9489.3&12017.2&12115.8 \\
Hornbaek&Denmark&1.50&8.1&200.0&7.0&166.9&1933.4&194.9&1007.3&2641.2&1003.8 \\
Gedser&Denmark&1.72&4.0&58.4&3.7&109.2&914.6&127.9&812.9&1668.7&781.2 \\
Baltra-B&Ecuador&0.80&76.9&122.9&127.6&3090.6&4669.4&5597.5&8256.7&10858.3&11005.8 \\
Santa Cruz&Ecuador&0.70&165.1&269.0&286.5&4832.6&6747.0&7910.7&9931.9&12317.2&12491.0 \\
La Libertad&Ecuador&0.85&185.8&282.8&285.1&6176.9&8310.8&9550.7&12931.6&14636.6&14967.8 \\
Acajutla-A&El Salvador&0.71&249.0&364.7&405.3&5935.2&8116.5&9234.2&11575.8&13611.7&13798.1 \\
Chuuk&Fd. St. Micronesia&0.63&93.9&153.5&136.5&5024.5&8055.6&9105.2&10345.1&12659.1&12949.8 \\
Kapingamarangi&Fd. St. Micronesia&0.60&370.7&489.7&608.8&8602.9&10565.8&11892.7&12922.2&14635.5&14914.7 \\
Pohnpei-B&Fd. St. Micronesia&0.63&327.7&451.9&453.8&7321.0&9974.5&11171.9&12247.0&14126.1&14473.3 \\
Yap-B&Fd. St. Micronesia&1.20&1.9&2.0&2.1&588.8&1063.0&1437.7&3989.9&6360.5&6126.2 \\
Suva-C&Fiji&0.63&676.9&845.7&873.2&12404.9&12672.6&14999.4&16038.2&16762.8&16397.8 \\
Noumea&France&0.50&893.1&1313.7&1252.1&10058.3&12285.0&13926.9&13310.3&15383.8&14906.0 \\
Brest&France&1.82&35.8&31.3&40.7&248.8&347.0&373.5&1065.9&1632.2&1257.6 \\
Marseille&France&0.87&15.6&506.2&21.7&955.6&4131.2&1427.3&4898.9&6665.8&6908.5 \\
Rikitea&French Polynesia&0.32&4672.5&6819.2&6884.8&15472.8&16486.4&16732.2&16129.4&16989.8&16665.1 \\
Papeete-B&French Polynesia&0.56&97.9&229.2&237.8&9207.3&11317.4&12597.7&13725.7&15603.6&15199.4 \\
Cuxhaven&Germany&3.48&2.2&2.2&2.1&5.9&16.4&6.7&117.5&101.0&93.2 \\
Malin Head&Ireland&1.51&6.4&7.7&4.7&149.2&267.7&146.4&845.7&755.3&783.9 \\
Hakodate&Japan&0.56&121.4&206.8&481.2&3463.5&5053.1&6687.4&7381.0&9884.3&10167.9 \\
Hamada&Japan&0.93&8.8&13.6&30.0&1588.7&2542.4&3719.3&6107.7&8879.3&9060.6 \\
Maisaka&Japan&1.45&1.2&1.3&1.4&85.9&99.1&157.6&629.8&965.9&1061.2 \\
Ishigaki&Japan&0.95&46.8&65.2&117.4&3098.8&4158.0&5698.0&8449.1&10967.0&11240.2 \\
Naha&Japan&0.94&33.7&56.0&128.7&2828.9&4040.8&5781.7&8093.6&10611.9&10994.4 \\
Toyama&Japan&0.61&286.0&502.4&1082.9&8545.6&10832.9&11987.9&12908.4&14962.6&14586.4 \\
Hosojima&Japan&1.17&3.3&4.1&6.0&405.0&596.9&920.6&2361.4&4121.0&4431.1 \\
Kushiro&Japan&0.67&5770.6&7626.4&8706.5&18242.0&18117.7&18238.4&18259.3&18172.3&18256.3 \\
Abashiri&Japan&0.76&101.5&259.1&308.5&3936.8&5903.6&7085.2&9504.7&11563.3&11859.1 \\
Mera&Japan&0.88&57.1&80.6&187.7&6567.0&8974.0&9846.8&13402.4&15161.9&14819.3 \\
Wakkanai&Japan&0.75&76.6&177.2&347.2&6017.5&8104.0&9465.9&12020.7&14193.7&13929.1 \\
Chichijima&Japan&0.85&35.2&139.0&220.5&5389.0&6648.4&8773.1&10891.1&12880.4&13115.3 \\
Nishinoomote&Japan&0.89&57.3&84.3&179.3&3296.9&4810.2&6170.3&8703.6&11442.7&11587.2 \\
Naze&Japan&1.00&40.0&55.9&125.3&3372.1&4686.5&6316.7&8986.2&11660.1&11951.1 \\
Hachinohe&Japan&0.66&652.2&903.8&1707.6&12559.9&14667.9&14962.3&16400.9&17081.9&16887.6 \\
Miyakejima&Japan&1.23&81.0&91.8&150.3&7494.5&9177.5&10099.1&15678.0&16699.8&16528.1 \\
Nakano Shima&Japan&1.08&13.9&20.0&48.7&2240.7&3089.9&4294.0&7494.9&10168.7&10345.9 \\
Ofunato&Japan&0.63&3985.9&5392.9&6782.0&18045.6&18091.0&17948.2&18250.3&18198.0&18151.4 \\
Nagasaki&Japan&0.96&91.8&123.2&177.3&2309.8&3395.0&4464.3&6755.1&9504.3&9602.6 \\
Aburatsu&Japan&1.18&5.5&6.9&10.9&658.9&977.1&1539.5&3773.4&6140.4&6470.8 \\
Kushimoto&Japan&0.82&210.8&280.6&507.9&8081.8&10405.9&11402.8&14079.5&15630.5&15450.2 \\
Cendering&Malaysia&1.10&20.5&27.1&32.8&973.1&1575.5&2104.6&4357.0&6496.5&6415.3 \\
Johor Baharu&Malaysia&0.93&114.7&144.8&172.0&2646.4&3864.4&4925.7&7320.0&9686.8&9712.5 \\
Kuantan&Malaysia&1.14&35.6&46.1&54.9&1182.9&1840.2&2439.9&4951.6&7159.9&7113.2 \\
Keling&Malaysia&0.74&158.1&201.3&246.1&3616.4&4995.7&6253.9&7926.8&10129.8&10155.5 \\
Lumut&Malaysia&0.84&123.2&163.9&202.9&2953.4&4220.8&5233.5&7256.8&9612.8&9513.9 \\
Kelang&Malaysia&1.33&52.4&61.5&72.0&755.2&1038.1&1377.7&3384.3&4913.7&4799.3 \\
Langkawi&Malaysia&0.91&144.8&189.4&231.5&2338.0&3451.5&4322.0&6369.0&8693.9&8597.9 \\
Penang&Malaysia&0.84&151.3&205.6&263.4&3388.3&4805.1&5845.1&7879.7&10228.2&10157.7 \\
Port Louis-C&Mauritius&0.46&548.2&945.1&1567.4&8746.7&11688.9&12436.9&12276.5&14568.0&14767.4 \\
Rodrigues&Mauritius&1.01&9.4&15.7&22.3&1720.5&2622.5&3365.4&6871.2&9698.1&9788.4 \\
Manzanillo-A&Mexico&0.63&686.2&1106.0&1163.5&14040.3&15038.2&15783.4&17026.6&17088.0&17308.2 \\
Ensenada&Mexico&0.70&338.2&507.0&543.0&5593.0&7748.8&8880.4&11298.3&13157.1&13458.9 \\
Salina Cruz&Mexico&0.69&190.2&295.0&320.5&7627.2&10012.8&11180.4&13303.4&14922.2&15125.8 \\
Acapulco-A, Gro.&Mexico&0.71&290.9&514.5&537.1&14612.1&15421.8&16170.1&17587.1&17458.4&17647.2 \\
Cabo San Lucas&Mexico&0.63&328.1&493.6&514.9&6135.7&8334.5&9398.6&11225.2&13194.9&13386.8 \\
Guaymas&Mexico&0.63&485.0&755.7&1056.3&12349.1&13855.1&14445.8&16311.3&16737.0&16976.6 \\
Saipan-B&N. Mariana Islands&1.32&1.4&1.5&1.5&308.5&641.1&588.2&2289.5&4034.0&4145.6 \\
Marsden Point&New Zealand&1.96&1.3&1.4&1.3&38.9&65.1&58.7&500.5&782.3&680.4 \\
Tauranga&New Zealand&0.58&254.9&922.0&696.5&6984.8&10869.3&11863.1&11422.6&14968.9&14995.1 \\
Taranaki&New Zealand&2.13&1.4&1.6&1.6&28.8&51.0&44.4&397.5&613.4&540.8 \\
Wellington&New Zealand&0.59&397.5&1303.0&962.0&9211.1&12970.2&13810.8&13415.5&16356.1&16440.7 \\
Tregde&Norway&0.99&12.7&35.2&10.6&449.5&925.4&537.7&2019.0&2231.5&2289.0 \\
Rorvik&Norway&1.72&1.2&1.6&1.2&39.8&91.0&49.0&295.8&296.5&266.9 \\
Ny-Alesund&Norway&0.76&1.9&136.8&0.1&52.5&179.2&45.0&463.2&207.5&287.2 \\
Vardo&Norway&1.23&5.7&51.5&10.6&184.1&599.4&307.5&853.4&1471.2&1104.1 \\
Balboa&Panama&1.33&85.2&101.9&109.9&762.7&1218.9&1353.8&3341.5&5591.5&5431.1 \\
Cristobal&Panama&0.35&4108.3&5087.1&6243.1&14803.5&16196.8&17285.0&15516.3&16766.0&17118.4 \\
Rabaul&Papua New Guinea&0.38&2202.9&2875.5&3499.5&12959.0&14228.3&15578.3&14822.4&16091.1&16294.8 \\
Lobos de Afuera&Peru&0.68&87.1&171.6&176.3&4281.2&6157.2&7186.3&9177.0&11646.8&11793.0 \\
Callao-B&Peru&0.62&26.8&71.3&87.6&3064.0&4690.2&5655.0&7598.6&10248.5&10467.0 \\
Legaspi&Philippines&0.64&689.0&1114.7&1221.4&12315.1&14158.3&15446.9&16071.0&16618.9&17111.8 \\
Manila&Philippines&0.84&1093.6&1505.3&1702.0&17367.2&17598.7&17960.9&18239.1&18087.3&18223.8 \\
Cascais&Portugal&1.04&65.7&57.0&89.2&948.0&1472.5&1966.8&4178.2&6325.2&6243.5 \\
Funchal-B&Portugal&0.68&665.3&600.7&934.1&6280.9&8180.3&9489.7&10988.8&12964.9&14017.0 \\
Kanton-B&Rep. of Kiribati&0.55&243.1&403.9&427.1&7791.4&9903.5&10701.4&12014.4&13890.0&14278.4 \\
Christmas-B&Rep. of Kiribati&0.48&911.0&1533.1&1574.7&12231.2&13488.8&14132.8&14586.3&15738.4&16052.9 \\
Majuro-A&Rep. of Marshall I&0.63&838.2&1069.4&1047.0&9263.5&11246.1&12550.0&13498.3&15041.0&15443.4 \\
Kwajalein&Rep. of Marshall I&0.54&1661.9&2085.6&2113.0&11845.5&13877.7&14831.2&14911.0&15887.9&16444.2 \\
Malakal-B&Republic of Belau&0.54&999.5&1290.4&1444.0&9808.3&13034.9&13861.9&13992.7&15584.5&15820.2 \\
Kaohsiung&Republic of China&0.80&14.5&24.3&42.1&1855.7&2710.9&3735.5&5372.6&7593.2&7765.6 \\
Keelung&Republic of China&1.04&7.1&9.1&15.8&1721.3&2594.8&3510.5&6428.6&8832.0&9048.7 \\
Luderitz&South Africa&0.61&458.6&574.9&704.3&7900.0&10339.8&11904.6&12388.8&14475.3&15083.9 \\
Saldahna Bay&South Africa&0.62&655.6&773.2&943.9&7636.9&10054.6&11266.4&11958.9&14120.9&14576.4 \\
Simon's Town&South Africa&0.72&269.4&339.8&425.1&6141.3&8623.2&10045.0&11775.4&14142.2&14630.1 \\
Port Nolloth&South Africa&0.69&352.7&434.7&467.1&6018.1&8424.1&9825.3&11258.6&13736.3&14277.5 \\
Port Elizabeth&South Africa&0.87&84.3&122.6&130.7&2446.5&3868.8&4964.6&7333.3&10394.5&10639.3 \\
La Coruna&Spain&1.17&112.8&97.3&125.1&1050.5&1502.6&1798.9&4397.5&6120.9&6073.2 \\
Ceuta&Spain&0.54&147.1&273.9&379.4&3718.6&5326.8&7051.5&8614.1&10996.1&11771.3 \\
Vigo&Spain&1.37&16.1&14.4&20.7&474.8&727.1&882.7&2764.4&4124.0&3864.9 \\
Stockholm&Sweden&0.98&3.1&1605.8&1.6&158.3&3866.6&162.5&536.3&100.1&605.1 \\
Goteborg-Torsh.&Sweden&1.43&3.2&45.5&2.7&113.9&787.9&122.7&681.4&1076.7&619.5 \\
Zanzibar&Tanzania&1.11&222.4&261.1&280.6&1660.9&2390.4&3059.0&5165.4&7698.0&7823.9 \\
Ko Lak&Thailand&1.05&117.8&146.6&165.9&2564.9&3808.8&4761.4&7812.3&10177.9&10187.1 \\
Stornoway&United Kingdom&1.50&28.8&26.5&23.6&245.7&346.0&256.4&1139.6&971.0&1080.2 \\
Lerwick&United Kingdom&1.02&17.0&25.3&13.6&448.5&682.4&470.2&1870.7&1791.5&1930.3 \\
Faraday&United Kingdom&1.19&1.9&2.2&2.0&13.9&18.0&37.4&579.7&2223.9&2099.3 \\
Gibraltar-A&United Kingdom&0.53&119.4&230.0&322.5&2770.5&4187.1&5636.8&6996.6&9516.1&10133.9 \\
Bermuda-B&United Kingdom&0.66&164.9&241.5&298.8&5611.6&7150.0&8131.5&10853.2&12681.1&12185.9 \\
Newlyn, Cornwall&United Kingdom&1.34&73.3&73.5&70.9&599.2&906.4&759.5&2386.8&2930.9&3136.2 \\
Seward-C, AK&USA&1.55&0.5&0.8&0.6&48.5&61.5&52.6&332.8&393.4&383.1 \\
Ketchikan, AK&USA&1.70&4.7&5.1&5.7&128.1&147.4&164.5&665.7&688.6&842.6 \\
Valdez, AK&USA&1.51&0.3&0.6&0.3&37.2&49.2&39.7&258.6&315.7&287.5 \\
Yakutat, AK&USA&1.37&0.0&0.0&0.0&14.9&28.3&16.7&146.3&141.1&144.4 \\
Seldovia, AK&USA&2.10&0.1&0.1&0.1&2.2&2.8&2.3&79.1&85.2&78.6 \\
Sitka, AK&USA&1.42&1.2&1.2&1.4&95.4&109.7&116.6&520.0&514.9&621.5 \\
Sand Point, AK&USA&1.28&13.1&14.4&16.9&444.4&514.4&683.8&2302.5&2505.8&3504.8 \\
Dutch Harbor-B, AK&USA&0.79&2.8&3.2&4.1&311.5&383.9&490.8&1150.2&1265.4&1710.8 \\
Cordova-B, AK&USA&1.64&8.2&11.4&10.1&311.3&406.3&406.9&2121.7&2236.7&2706.9 \\
Kodiak Isl., AK&USA&1.52&0.3&0.3&0.3&6.1&10.3&7.4&125.8&122.3&126.1 \\
Adak, AK&USA&0.96&6.0&8.6&9.0&454.9&598.3&760.2&1785.3&1974.0&2752.1 \\
San Francisco, CA&USA&0.89&20.7&39.6&42.1&1823.3&2758.9&3789.1&7063.3&9574.6&9629.9 \\
San Diego, CA&USA&0.77&249.4&378.4&409.7&4589.5&6580.3&7673.4&10697.9&12691.8&13011.9 \\
Los Angeles, CA&USA&0.72&168.0&285.6&303.7&3092.5&4691.0&5715.1&7958.2&10421.5&10575.5 \\
Crescent City, CA&USA&1.10&4.7&6.9&8.4&329.2&400.4&604.8&1600.8&2592.4&2418.6 \\
Monterey, CA&USA&0.81&49.0&92.6&105.3&2320.4&3533.9&4617.0&7377.1&9852.3&9937.1 \\
Port San Luis, CA&USA&0.82&35.1&71.1&79.3&1693.0&2547.6&3502.1&5824.4&8377.4&8399.8 \\
Santa Monica, CA&USA&0.84&54.9&112.8&116.2&2190.4&3355.7&4275.6&7059.2&9563.1&9654.0 \\
La Jolla, CA&USA&0.72&327.4&492.6&533.5&5515.6&7712.7&8809.0&11490.9&13299.8&13633.6 \\
New London, CT&USA&1.68&2.9&3.2&3.6&155.6&300.8&281.9&1031.5&2694.6&1706.5 \\
Fernandina Beach, FL&USA&1.20&13.4&15.3&19.2&669.9&1379.4&1489.1&3993.0&6869.0&6668.8 \\
St. Petersburg, FL&USA&1.54&2.1&2.3&2.3&190.1&365.1&339.0&1715.4&3620.5&3023.4 \\
Pensacola, FL&USA&2.23&1.3&1.3&1.4&32.0&50.5&45.9&375.1&688.0&453.8 \\
Mayport, FL&USA&0.79&93.2&121.0&172.2&3408.7&6302.9&7086.3&9440.2&12136.4&13088.3 \\
Limetree Bay, FL&USA&0.88&2.0&2.7&2.4&1333.9&2503.3&2629.5&6559.3&9187.5&9575.8 \\
Key West, FL&USA&0.67&91.1&175.2&197.6&6284.3&8905.2&10664.8&12006.7&14008.6&14614.5 \\
Fort Pulaski, GA&USA&0.98&56.7&67.4&101.0&2352.9&4579.8&5141.2&8604.6&11498.5&12311.0 \\
Hilo, HI&USA&0.55&1018.3&1286.5&1778.9&13772.9&14962.4&15590.6&15977.4&16863.3&16750.6 \\
French Frigate, HI&USA&0.44&1585.8&2755.6&2907.1&13046.3&14904.3&15514.9&14740.9&16107.9&16072.8 \\
Kahului, HI&USA&0.45&2203.3&2720.3&3559.6&14790.3&15690.8&16221.3&16039.3&16914.9&16808.8 \\
Mokuoloe, HI&USA&0.41&2336.8&2972.8&3796.2&14345.2&15328.8&15919.7&15418.1&16573.7&16502.0 \\
Honolulu-B, HI&USA&0.41&2288.7&2920.3&3732.4&14323.4&15312.1&15908.4&15407.6&16565.8&16496.3 \\
Nawiliwili, HI&USA&0.53&408.5&567.5&740.0&9834.7&11852.7&12691.7&13099.8&14987.4&14967.3 \\
Grand Isle, LA&USA&3.70&1.3&1.3&1.4&2.0&2.4&2.2&213.1&353.1&237.2 \\
Woods Hole, MA&USA&1.31&8.1&10.6&11.6&525.7&1263.0&1148.7&2705.4&6116.4&5020.3 \\
Nantucket, MA&USA&2.03&1.7&1.7&1.8&79.3&138.7&130.0&639.6&1478.5&902.8 \\
Boston, MA&USA&1.41&9.3&12.0&13.7&452.7&952.4&952.8&2193.9&5046.5&3868.8 \\
Portland, ME&USA&1.16&26.1&38.4&42.5&772.0&1748.1&1693.5&2849.2&6214.0&5109.9 \\
Eastport, ME&USA&1.51&26.1&39.8&45.0&473.7&1082.2&1013.1&1806.9&4375.2&2922.4 \\
Duck Pier, NC&USA&1.17&25.8&42.2&49.3&1311.4&3153.4&3397.8&6650.5&10067.6&10437.7 \\
Wilmington, NC&USA&0.90&11.3&58.9&40.6&1509.8&3606.1&4054.4&6768.7&10082.3&10484.1 \\
Atlantic City, NJ&USA&1.30&14.1&21.7&24.7&927.0&2332.7&2314.9&4953.8&8873.5&8760.8 \\
Cape May, NJ&USA&1.17&24.4&33.5&41.2&1210.4&3345.0&3225.2&6092.6&9903.4&10465.0 \\
Montauk, NY&USA&1.27&12.8&17.1&19.3&666.4&1504.0&1488.6&3391.5&7137.6&6538.6 \\
New York, NY&USA&1.80&2.6&2.8&3.1&138.0&259.9&245.7&967.6&2498.4&1532.1 \\
Charleston, OR&USA&1.09&20.8&27.0&33.3&711.1&899.9&1379.6&3313.8&5117.7&4931.5 \\
South Beach, OR&USA&1.23&18.5&23.8&28.1&696.8&869.4&1298.7&3578.9&5487.9&5244.4 \\
Astoria, OR&USA&1.15&7.0&9.7&11.9&335.0&408.3&586.2&1583.4&2581.5&2326.1 \\
Newport, RI&USA&1.18&13.8&21.9&22.7&860.9&2096.2&1980.9&3811.0&7640.1&7032.3 \\
Charleston, SC&USA&1.16&8.0&11.5&13.3&865.5&1779.7&2087.7&5213.8&8353.3&8455.6 \\
Galveston (Pier 21), TX&USA&1.85&3.0&3.2&3.2&216.1&448.7&417.7&3109.2&5934.6&5325.5 \\
Rockport, TX&USA&1.18&5.7&13.6&7.1&1697.3&3931.9&4450.4&10313.5&13021.6&13464.6 \\
Port Isabel, TX&USA&1.36&2.3&2.5&2.5&358.2&803.8&837.9&3580.4&6461.5&5865.2 \\
Galveston (P. Pier), TX&USA&2.62&1.7&1.7&1.7&37.5&63.8&61.2&609.4&1213.0&856.4 \\
Chesapeake BBT, VA&USA&1.53&6.5&7.8&9.1&386.1&881.5&922.6&3030.7&6147.3&5476.5 \\
Neah Bay, WA&USA&1.24&2.4&3.8&5.2&168.4&201.4&265.5&807.5&1194.2&1053.1 \\
Willapa Bay, WA&USA&1.74&2.7&3.2&3.6&113.1&127.9&177.7&747.7&1114.7&994.0 \\
Lewes, DE&USA&1.59&4.2&4.8&5.5&256.8&652.8&605.9&1951.0&4669.4&3575.8 \\
Apra Harbor, Guam&USA Trust&0.40&2125.7&2788.0&3372.7&13523.0&14744.7&16219.8&15396.9&16164.8&16764.4 \\
Wake&USA Trust&1.00&4.2&29.7&6.8&1898.4&3156.2&3458.0&6520.4&9154.8&9473.0 \\
Johnston&USA Trust&0.74&71.0&120.9&102.5&4510.1&6877.5&7501.6&9348.6&11785.2&12101.8 \\
Midway&USA Trust&0.88&13.4&46.8&39.9&2185.3&3755.4&5136.5&7521.3&9986.8&10271.9 \\
Pago Pago&USA Trust&0.45&2208.9&3382.2&3665.0&15823.0&16830.0&16978.0&16923.8&17456.9&17394.1 \\
Charlotte Amalie, VI&USA Trust&0.64&56.0&76.3&85.1&4840.3&7220.3&8019.9&10661.5&12853.7&13519.8 \\
San Juan, PR&USA Trust&0.57&171.0&214.1&257.2&7621.6&9766.5&11001.1&12471.7&14218.7&14985.5 \\
Magueyes Island, PR&USA Trust&0.56&116.0&168.7&183.6&7002.7&9212.3&10381.2&11933.0&13864.4&14527.4 \\
\hline
\end{longtable}}
\label{Stab:af_table_rp100}
\end{landscape}

\begin{landscape}
\centering
\setlength{\tabcolsep}{1pt}
{\small
\begin{longtable}{llcccc|ccc|ccc}
\caption{Expected flood amplification factors (AF) for the 500-yr flood for 2050, 2100, and 2150 under 1.5$^\circ$C, 2.0$^\circ$C, and 2.5$^\circ$C global mean surface  temperature stabilization scenarios.} \\
& & \multicolumn{9}{c}{500-yr Flood} \\
& & & \multicolumn{3}{c}{2050} & \multicolumn{3}{c}{2100} & \multicolumn{3}{c}{2150} \\ \cmidrule(lr){4-6}  \cmidrule(lr){7-9} \cmidrule(lr){10-12} 
Site & Region & Historical Height (m above MHHW)&AF 1.5$^\circ$C & AF 2.0$^\circ$C & AF 2.5$^\circ$C&AF 1.5$^\circ$C & AF 2.0$^\circ$C & AF 2.5$^\circ$C&AF 1.5$^\circ$C & AF 2.0$^\circ$C & AF 2.5$^\circ$C \\
\hline
\endfirsthead
\multicolumn{9}{c}
{{\bfseries \tablename\ \thetable{} -- continued from previous page}} \\
& & \multicolumn{9}{c}{500-yr Flood} \\
& & & \multicolumn{3}{c}{2050} & \multicolumn{3}{c}{2100} & \multicolumn{3}{c}{2150} \\ \cmidrule(lr){4-6}  \cmidrule(lr){7-9} \cmidrule(lr){10-12} 
Site & Region & Historical Height (m above MHHW)&AF 1.5$^\circ$C & AF 2.0$^\circ$C & AF 2.5$^\circ$C&AF 1.5$^\circ$C & AF 2.0$^\circ$C & AF 2.5$^\circ$C&AF 1.5$^\circ$C & AF 2.0$^\circ$C & AF 2.5$^\circ$C \\
\hline
\endhead
\hline
\multicolumn{9}{c}{{Continued on next page}} \\
\endfoot
\endlastfoot
Buenos Aires&Argentina&4.05&1.4&1.4&1.4&2.1&2.5&2.7&114.6&180.3&138.1 \\
Fort Denison&Australia&0.81&140.5&429.6&323.5&8649.3&16123.9&19873.8&30225.0&48052.3&49762.5 \\
Bundaberg&Australia&1.61&2.7&3.6&3.3&665.9&1002.4&1159.0&6239.1&11760.0&11111.4 \\
Brisbane&Australia&0.89&104.4&261.1&201.9&8448.1&15684.0&17844.6&30889.7&48056.7&46981.0 \\
Spring Bay&Australia&0.84&60.0&176.3&244.3&7958.9&18449.5&19846.2&33203.7&50567.9&50639.5 \\
Townsville&Australia&1.49&33.5&61.3&51.7&1626.8&2518.0&3219.9&10861.1&19537.4&18869.4 \\
Broome&Australia&2.50&85.6&107.6&126.3&986.1&1363.2&1586.5&4122.9&6203.4&5810.1 \\
Cocos&Australia&0.65&1958.0&2546.6&4139.6&43909.2&56561.1&62643.3&69259.0&76339.7&78271.0 \\
Darwin&Australia&1.83&40.7&59.7&70.5&1609.9&2322.3&2771.1&8357.8&13279.4&12654.3 \\
Esperance&Australia&0.85&413.0&543.7&734.4&13232.3&20200.7&25945.4&40484.4&54894.6&55731.2 \\
Fremantle&Australia&1.00&60.3&96.4&141.3&6270.9&10620.5&13973.1&29603.4&45299.6&45251.7 \\
Cananeia&Brazil&1.79&3.4&4.0&3.9&495.9&1019.3&1007.0&7657.8&16509.4&13643.6 \\
Ilha Fiscal, RJ&Brazil&1.24&13.3&22.0&19.1&2160.3&4752.8&4854.5&17847.6&30823.6&29175.2 \\
Victoria, BC&Canada&1.14&19.3&28.1&37.7&2009.5&2498.2&3477.7&10122.1&16606.5&14999.5 \\
Prince Rupert&Canada&1.84&18.2&20.7&22.4&752.0&870.2&974.1&4239.0&4457.0&5507.5 \\
Tofino&Canada&1.39&3.7&5.3&7.5&661.0&775.3&981.2&3594.3&5176.1&4638.8 \\
St. John's-A&Canada&1.20&27.2&56.6&58.5&2832.2&7481.8&7033.6&11938.1&29799.7&19925.7 \\
Halifax&Canada&1.45&9.5&14.0&14.9&1852.9&4730.2&3983.4&10299.0&26082.2&17564.1 \\
Churchill&Canada&1.94&0.2&0.3&0.3&1.6&23.4&5.0&282.3&276.2&259.8 \\
Puerto Montt&Chile&1.76&46.1&65.4&82.6&774.3&1159.8&1435.0&4023.0&6896.6&6573.1 \\
Juan Fernandez-B&Chile&0.69&115.8&210.1&264.1&11187.0&17942.3&23604.7&35908.8&48786.8&49458.1 \\
Antofagasta&Chile&0.71&20.7&61.1&72.1&5740.8&10700.9&13566.5&23918.9&37939.5&38056.6 \\
Easter-C&Chile&1.29&2.3&2.5&2.6&970.3&1351.1&1694.8&9936.0&16619.2&18445.9 \\
Valparaiso&Chile&0.66&69.6&160.1&180.7&5901.1&11314.7&14270.5&23035.9&36904.8&37161.8 \\
Xiamen&China&1.93&8.4&10.7&14.2&601.3&740.5&1133.8&4995.8&8171.5&8496.3 \\
Buenaventura&Colombia&1.30&81.9&150.2&148.8&3598.2&5729.0&6624.4&17997.7&29581.2&27909.2 \\
Tumaco&Colombia&1.11&22.6&51.3&50.9&2194.3&3645.2&4291.0&11067.3&19954.9&18413.6 \\
Cartagena&Colombia&0.34&77164.9&69958.0&78217.0&90984.6&90137.6&91235.8&90812.4&89782.8&91017.1 \\
Penrhyn&Cook Islands&0.75&50.8&177.5&75.4&18978.2&26800.6&32871.6&48939.6&60896.9&62077.4 \\
Quepos-A&Costa Rica&0.81&960.8&1310.0&1381.3&18173.4&27379.4&31975.9&45697.4&58523.8&58995.0 \\
Hornbaek&Denmark&1.65&12.4&653.9&10.2&541.1&7453.4&611.9&3749.3&11200.9&3511.6 \\
Gedser&Denmark&2.02&4.1&116.0&3.7&256.4&2555.1&282.6&2308.0&5303.9&2100.0 \\
Baltra-B&Ecuador&0.87&180.7&300.0&311.6&11000.5&17367.9&21171.3&35614.1&49127.9&49572.8 \\
Santa Cruz&Ecuador&0.73&593.1&1007.8&1059.3&21138.3&30282.1&35746.0&47025.4&59364.5&60190.0 \\
La Libertad&Ecuador&0.92&452.8&719.0&725.7&22430.5&32398.4&37712.2&58035.9&68550.3&69871.1 \\
Acajutla-A&El Salvador&0.77&622.5&989.9&1089.5&22499.7&32654.1&37598.3&52280.4&63639.0&64561.5 \\
Chuuk&Fd. St. Micronesia&1.01&3.0&4.8&3.2&4159.7&9376.6&10166.9&24077.9&35763.0&35998.8 \\
Kapingamarangi&Fd. St. Micronesia&0.67&765.7&1032.3&1298.7&32780.5&43647.5&49632.2&58552.4&68626.0&69991.1 \\
Pohnpei-B&Fd. St. Micronesia&0.72&556.1&786.6&761.2&25024.7&38395.8&43866.7&53733.5&64683.1&66171.7 \\
Yap-B&Fd. St. Micronesia&2.64&1.2&1.3&1.2&29.7&137.2&148.4&1287.5&1900.4&1612.4 \\
Suva-C&Fiji&0.73&1006.2&1301.6&1352.0&46453.7&50845.0&62450.7&73297.6&79385.0&77276.1 \\
Noumea&France&0.54&2716.1&3967.5&3795.2&43692.9&55365.7&64369.5&62814.9&74443.8&71836.0 \\
Brest&France&1.87&115.4&97.7&134.9&1041.6&1469.7&1587.8&4800.2&7394.6&5620.2 \\
Marseille&France&0.98&22.8&1545.3&32.2&2845.9&16949.3&4078.8&17836.1&31071.6&25817.0 \\
Rikitea&French Polynesia&0.34&17769.8&27627.1&27724.8&74777.4&80591.4&82129.7&79346.0&84219.9&82615.2 \\
Papeete-B&French Polynesia&0.94&2.9&3.9&3.8&7289.1&12427.6&14575.5&36066.2&50044.1&48578.7 \\
Cuxhaven&Germany&4.09&2.0&2.1&2.0&5.7&14.1&6.5&294.0&248.7&222.8 \\
Malin Head&Ireland&1.68&7.0&10.0&4.7&463.4&861.2&429.7&3062.6&2695.1&2733.3 \\
Hakodate&Japan&0.61&306.9&552.7&1445.1&13464.2&19964.6&27948.1&32966.9&45433.9&47047.5 \\
Hamada&Japan&1.30&2.8&3.3&4.5&1618.3&2299.6&3735.2&12026.5&20931.3&22583.2 \\
Maisaka&Japan&2.48&1.1&1.1&1.2&1.9&3.6&14.3&749.8&944.4&912.8 \\
Ishigaki&Japan&1.09&57.3&83.1&154.0&8209.6&11067.3&16926.9&31923.0&44837.0&46038.4 \\
Naha&Japan&1.17&13.7&22.7&56.1&4885.3&7073.3&12090.8&24776.0&37374.6&38516.4 \\
Toyama&Japan&0.75&196.6&415.9&1216.8&23548.6&33811.3&41563.2&51716.9&64782.3&63765.6 \\
Hosojima&Japan&1.56&2.0&2.3&2.7&589.9&748.6&1109.7&4699.7&7966.2&8917.5 \\
Kushiro&Japan&0.78&8512.9&14900.2&19000.6&90460.1&88966.1&90583.9&91285.0&90484.2&91212.0 \\
Abashiri&Japan&0.85&162.1&513.4&610.4&12710.8&20272.1&25672.6&39882.6&50971.5&52256.1 \\
Mera&Japan&1.20&9.5&11.6&22.8&7026.7&11955.7&16064.0&38523.0&52089.8&53518.6 \\
Wakkanai&Japan&0.99&15.1&43.1&106.8&9517.2&13334.5&20652.1&39038.1&51548.3&52521.4 \\
Chichijima&Japan&1.19&5.4&19.4&22.3&7002.6&9048.4&14424.1&29332.4&41111.9&42764.4 \\
Nishinoomote&Japan&1.04&46.7&73.9&181.8&8063.7&12476.5&17797.7&31876.0&45726.1&47144.8 \\
Naze&Japan&1.29&10.1&13.6&30.6&4535.2&6523.4&10243.3&24213.2&37680.9&38863.4 \\
Hachinohe&Japan&0.82&455.1&641.7&1679.5&38015.3&49823.7&55645.8&70360.0&78037.2&77295.2 \\
Miyakejima&Japan&1.68&8.7&9.5&14.5&5691.5&7779.2&11104.7&42015.4&55128.9&57206.1 \\
Nakano Shima&Japan&1.55&2.8&3.1&4.3&1595.5&2033.6&3230.7&12922.6&21883.8&23538.5 \\
Ofunato&Japan&0.74&5692.6&8106.9&14596.3&85635.4&88070.9&86722.2&91097.5&90633.4&90151.2 \\
Nagasaki&Japan&1.05&155.0&230.5&380.2&7696.2&11484.2&15893.4&27698.3&40991.7&41803.5 \\
Aburatsu&Japan&1.53&3.1&3.7&4.9&941.1&1229.0&1932.8&7726.6&13828.8&15228.5 \\
Kushimoto&Japan&0.92&365.5&502.3&973.7&26794.1&37762.8&44224.2&61662.8&72109.4&71444.6 \\
Cendering&Malaysia&1.29&12.8&17.3&22.4&2219.1&3586.5&4902.8&14357.5&22999.0&22493.2 \\
Johor Baharu&Malaysia&1.00&266.2&353.6&435.4&9829.7&14816.3&19296.0&32162.4&43810.1&43972.5 \\
Kuantan&Malaysia&1.29&33.8&45.8&58.8&3175.4&5016.7&6817.2&18037.4&27665.7&27214.0 \\
Keling&Malaysia&0.80&390.5&514.9&642.1&13908.9&19823.4&25568.4&35632.2&46594.1&46660.6 \\
Lumut&Malaysia&0.88&409.8&558.8&700.9&12477.2&18077.1&22858.0&33750.8&45350.8&44870.4 \\
Kelang&Malaysia&1.40&130.7&162.4&196.4&2893.0&3988.5&5357.5&14546.5&21481.1&20934.8 \\
Langkawi&Malaysia&0.97&385.7&542.3&685.2&9124.7&13723.9&17524.9&28361.7&39755.8&39224.2 \\
Penang&Malaysia&0.94&225.7&349.7&480.9&11125.0&16447.1&20902.1&32935.2&44464.6&44063.9 \\
Port Louis-C&Mauritius&0.49&1848.3&3257.4&5610.2&38976.8&53603.6&57972.4&58577.8&70815.7&71760.1 \\
Rodrigues&Mauritius&1.42&2.8&3.2&3.9&1623.6&2260.9&2897.7&12859.2&21911.1&21991.9 \\
Manzanillo-A&Mexico&0.72&1131.0&1937.6&2033.4&55303.8&64018.1&68917.4&79934.6&82282.4&83662.1 \\
Ensenada&Mexico&0.73&1267.7&1954.2&2094.0&24514.4&34709.0&40349.5&53663.9&63585.7&65056.7 \\
Salina Cruz&Mexico&0.89&50.1&106.9&101.5&14457.5&23775.3&27655.5&48136.9&60529.7&61396.3 \\
Acapulco-A, Gro.&Mexico&0.90&118.8&271.7&237.9&39725.2&50443.3&56119.2&77573.4&80842.2&82411.1 \\
Cabo San Lucas&Mexico&0.64&1468.5&2228.6&2324.3&29384.7&40279.6&45597.8&55212.5&65245.7&66217.7 \\
Guaymas&Mexico&0.72&807.9&1353.9&1916.9&45643.1&55248.7&60049.0&75171.2&79552.5&80992.7 \\
Saipan-B&N. Mariana Islands&3.85&1.1&1.1&1.1&1.2&1.3&1.3&282.0&385.1&316.3 \\
Marsden Point&New Zealand&5.38&1.1&1.1&1.1&1.2&1.2&1.2&1.3&21.3&1.4 \\
Tauranga&New Zealand&0.63&667.2&2631.1&1973.5&27072.0&46343.4&51139.3&52363.9&71123.0&71050.7 \\
Taranaki&New Zealand&3.99&1.2&1.2&1.2&1.4&1.5&1.6&183.3&200.9&173.9 \\
Wellington&New Zealand&0.65&857.0&3345.0&2400.7&34912.3&54956.9&59242.1&61140.5&78363.7&78370.3 \\
Tregde&Norway&1.18&9.3&28.4&7.7&1168.3&2385.1&1331.5&6488.8&6766.6&6707.0 \\
Rorvik&Norway&2.39&1.1&1.2&1.1&11.5&131.0&14.3&643.0&591.0&520.2 \\
Ny-Alesund&Norway&0.82&4.9&573.3&0.1&229.0&802.7&192.6&2165.0&961.3&1338.6 \\
Vardo&Norway&1.37&6.6&93.7&13.9&594.2&1953.8&972.7&3144.4&5343.2&3985.6 \\
Balboa&Panama&1.38&263.7&335.7&362.2&3156.8&5072.9&5620.0&14822.8&25219.9&24356.5 \\
Cristobal&Panama&0.40&9278.2&13596.8&16804.0&65421.6&75619.8&82205.6&73819.6&81834.0&83668.5 \\
Rabaul&Papua New Guinea&0.38&11014.3&14377.8&17497.4&64795.2&71141.3&77891.7&74111.8&80455.4&81473.9 \\
Lobos de Afuera&Peru&0.72&276.2&573.8&575.1&17694.5&26444.9&31047.5&42506.6&55264.3&55909.7 \\
Callao-B&Peru&0.80&8.2&21.4&24.4&5680.1&9493.2&11853.5&24474.1&37730.3&37804.3 \\
Legaspi&Philippines&0.77&672.4&1280.7&1393.6&39288.3&52445.8&60979.7&70550.7&77203.3&79735.2 \\
Manila&Philippines&0.98&1466.0&2092.1&2382.9&71211.7&78505.8&83330.9&90527.9&89315.4&90575.8 \\
Cascais&Portugal&1.11&157.1&133.1&230.1&3470.9&5448.2&7299.2&17329.6&27031.0&26251.7 \\
Funchal-B&Portugal&0.71&2580.3&2331.7&3677.8&27776.2&36911.4&43573.5&52316.2&62627.8&67887.2 \\
Kanton-B&Rep. of Kiribati&0.66&300.1&460.6&522.5&23931.7&34077.5&38042.5&50457.3&61644.6&63473.8 \\
Christmas-B&Rep. of Kiribati&0.52&2621.6&4558.7&4627.9&54577.5&61850.0&65740.1&69963.3&76608.6&78094.6 \\
Majuro-A&Rep. of Marshall I&0.65&3486.6&4463.4&4341.8&43283.8&53465.3&60070.9&65882.6&73996.2&76056.2 \\
Kwajalein&Rep. of Marshall I&0.55&7535.2&9514.0&9580.9&57710.2&68122.4&73093.5&73831.2&78943.6&81762.3 \\
Malakal-B&Republic of Belau&0.56&4020.4&5228.4&5829.8&46135.4&62482.2&67019.1&68355.4&76810.0&77964.2 \\
Kaohsiung&Republic of China&1.07&3.5&4.8&7.6&2745.8&4086.2&6102.2&14794.9&23462.8&23993.9 \\
Keelung&Republic of China&1.43&3.4&3.8&4.9&1656.4&2444.9&3500.2&13853.1&22131.6&23450.5 \\
Luderitz&South Africa&0.64&1614.5&2063.5&2531.6&34663.2&46826.7&54805.1&58991.6&70359.4&73323.0 \\
Saldahna Bay&South Africa&0.63&2972.8&3519.5&4297.4&36671.4&48734.4&54863.8&58856.5&69929.0&72186.4 \\
Simon's Town&South Africa&0.80&495.8&681.2&872.5&20788.9&31289.8&38108.3&51010.2&64791.4&66982.6 \\
Port Nolloth&South Africa&0.72&1312.4&1631.3&1750.4&26278.7&37710.8&44505.3&53500.3&66525.6&69169.3 \\
Port Elizabeth&South Africa&0.94&188.3&296.3&315.1&8572.9&13846.3&18411.2&31228.4&46492.7&47493.2 \\
La Coruna&Spain&1.22&364.1&306.5&413.6&4292.4&6156.7&7399.9&19453.0&27394.6&26973.1 \\
Ceuta&Spain&0.60&339.5&731.1&1008.9&12885.0&19906.2&26779.3&37106.1&49493.7&52859.9 \\
Vigo&Spain&1.63&8.1&7.5&9.8&884.8&1383.2&1635.4&7279.0&11009.7&9524.1 \\
Stockholm&Sweden&1.14&3.1&6292.0&1.6&526.3&16766.7&528.8&1993.9&418.2&2212.6 \\
Goteborg-Torsh.&Sweden&1.69&2.7&51.3&2.2&296.8&2278.5&301.7&2120.0&3355.7&1873.8 \\
Zanzibar&Tanzania&1.13&960.4&1144.4&1236.9&7677.9&11082.1&14232.0&24700.0&37153.7&37779.2 \\
Ko Lak&Thailand&1.13&276.0&360.1&418.6&9199.4&14085.2&18055.8&33798.1&45554.8&45542.2 \\
Stornoway&United Kingdom&1.56&81.1&74.4&62.4&1003.9&1428.9&1030.0&5066.7&4250.9&4684.9 \\
Lerwick&United Kingdom&1.17&17.6&30.9&13.2&1336.2&2017.8&1346.9&6700.9&6205.4&6569.6 \\
Faraday&United Kingdom&1.77&1.3&1.4&1.3&1.8&2.0&2.6&279.2&1830.0&1590.9 \\
Gibraltar-A&United Kingdom&0.63&116.6&408.4&543.8&7593.1&12690.0&16945.0&26366.4&38980.4&41214.7 \\
Bermuda-B&United Kingdom&0.79&177.3&347.4&366.9&15488.1&21459.5&25739.3&43454.5&52731.1&49505.4 \\
Newlyn, Cornwall&United Kingdom&1.39&231.9&233.1&223.2&2478.0&3797.5&3131.7&10573.4&12929.0&13739.5 \\
Seward-C, AK&USA&1.78&0.6&0.8&0.7&133.1&186.8&135.2&1223.6&1455.2&1350.9 \\
Ketchikan, AK&USA&1.81&6.4&6.5&7.8&446.7&522.4&559.6&2715.9&2786.7&3318.2 \\
Valdez, AK&USA&1.69&0.3&0.6&0.4&88.2&169.3&92.6&1024.5&1258.7&1116.4 \\
Yakutat, AK&USA&1.42&0.0&0.0&0.0&57.4&109.3&64.6&694.6&660.5&677.3 \\
Seldovia, AK&USA&2.37&0.2&0.2&0.2&3.9&5.1&4.0&306.7&343.9&300.5 \\
Sitka, AK&USA&1.55&1.5&1.4&1.7&338.6&395.8&400.5&2107.0&2082.7&2409.4 \\
Sand Point, AK&USA&1.37&24.6&27.6&33.7&1609.8&1866.0&2447.9&9292.2&9977.8&14117.1 \\
Dutch Harbor-B, AK&USA&0.82&8.9&9.7&12.8&1408.5&1743.0&2193.2&5431.6&5939.9&7931.9 \\
Cordova-B, AK&USA&1.92&4.7&6.5&5.6&623.3&796.4&768.0&5696.0&6027.9&7050.1 \\
Kodiak Isl., AK&USA&2.09&0.5&0.5&0.5&1.0&1.8&1.2&356.5&343.7&322.7 \\
Adak, AK&USA&1.11&5.2&8.4&8.2&1383.2&1815.2&2132.2&6278.6&6874.7&9347.5 \\
San Francisco, CA&USA&1.06&15.5&29.9&31.8&4100.7&5853.1&8906.2&22800.8&35021.0&34393.6 \\
San Diego, CA&USA&0.82&685.6&1184.8&1254.6&18121.6&26799.8&32060.9&48584.5&59534.8&60938.9 \\
Los Angeles, CA&USA&0.76&500.0&966.2&996.9&12694.3&19522.0&24339.8&36256.4&48891.0&49459.8 \\
Crescent City, CA&USA&1.20&7.3&11.5&14.4&1169.1&1400.0&2108.9&6403.9&10043.0&9338.2 \\
Monterey, CA&USA&0.90&75.9&161.7&185.9&7473.5&11305.8&15807.4&29701.9&42331.8&42333.4 \\
Port San Luis, CA&USA&0.91&51.8&122.0&136.0&5535.8&8139.4&11871.8&23099.6&35399.1&34987.0 \\
Santa Monica, CA&USA&0.96&52.0&135.9&135.7&6182.5&9383.3&12858.6&26153.0&38887.2&38665.6 \\
La Jolla, CA&USA&0.76&1083.4&1715.4&1865.3&22921.9&33054.4&38521.9&53473.5&63448.4&65032.2 \\
New London, CT&USA&2.32&2.1&2.1&2.4&180.9&251.4&291.2&1676.3&3542.3&2172.2 \\
Fernandina Beach, FL&USA&1.47&7.1&8.4&10.1&1227.4&2412.2&2486.5&9862.5&19474.0&16696.3 \\
St. Petersburg, FL&USA&2.61&1.4&1.5&1.5&49.6&156.3&142.7&1325.6&2260.5&1578.4 \\
Pensacola, FL&USA&4.97&1.1&1.1&1.1&1.2&1.3&1.3&121.9&167.6&110.6 \\
Mayport, FL&USA&0.95&57.6&93.4&124.3&7375.9&15849.8&17493.3&33130.5&48435.1&50600.1 \\
Limetree Bay, FL&USA&2.46&1.2&1.1&1.2&108.3&139.3&138.7&1367.8&2040.6&1641.1 \\
Key West, FL&USA&0.96&7.9&31.9&16.6&6715.3&14048.9&15782.9&33432.6&48509.5&49892.8 \\
Fort Pulaski, GA&USA&1.17&33.9&43.5&65.2&4763.4&9952.9&11180.8&27439.7&43026.5&43911.6 \\
Hilo, HI&USA&0.67&1101.0&1398.0&2054.7&48909.9&58668.3&64368.1&71676.9&78764.7&78545.2 \\
French Frigate, HI&USA&0.47&5112.1&9501.6&9810.6&60720.3&71033.6&74442.4&71625.4&79126.1&78932.7 \\
Kahului, HI&USA&0.52&4370.1&5496.7&7507.0&63429.8&70423.9&74299.0&75574.3&81667.8&81311.8 \\
Mokuoloe, HI&USA&0.44&7846.7&10190.2&13258.2&67323.2&73246.4&76634.2&75038.4&81544.9&81232.8 \\
Honolulu-B, HI&USA&0.45&6664.0&8728.3&11408.8&65557.3&71928.3&75504.8&74255.5&81007.5&80723.4 \\
Nawiliwili, HI&USA&0.71&224.1&262.8&392.7&23574.0&33271.0&38319.2&49882.4&62393.3&62933.5 \\
Grand Isle, LA&USA&10.00&1.3&1.2&1.3&1.4&1.4&1.4&1.5&1.6&1.6 \\
Woods Hole, MA&USA&1.65&5.2&6.5&7.2&890.1&1945.4&1658.5&5841.4&15392.8&9538.4 \\
Nantucket, MA&USA&3.74&1.2&1.3&1.3&1.7&1.9&2.0&390.1&651.2&417.8 \\
Boston, MA&USA&1.75&5.1&6.2&7.0&782.7&1499.7&1401.0&5036.4&12481.3&7633.6 \\
Portland, ME&USA&1.33&26.3&41.8&44.3&2054.5&4493.9&4260.2&9109.6&22304.1&15760.2 \\
Eastport, ME&USA&1.72&19.3&33.7&34.5&1154.2&2723.6&2423.1&5694.6&14587.3&8648.3 \\
Duck Pier, NC&USA&1.40&17.9&29.9&35.3&2492.5&5948.8&6505.1&18241.9&34590.0&31860.1 \\
Wilmington, NC&USA&1.24&3.9&7.7&6.0&1782.8&4022.9&4379.2&13403.3&27017.5&23543.8 \\
Atlantic City, NJ&USA&1.62&8.0&10.9&12.6&1407.0&3299.1&3205.9&10439.6&24855.3&19048.4 \\
Cape May, NJ&USA&1.43&14.5&19.6&24.3&2118.9&5730.6&5395.8&14979.9&32360.6&28499.7 \\
Montauk, NY&USA&1.53&10.6&13.7&16.0&1350.4&2788.1&2665.4&8412.9&21327.4&15755.3 \\
New York, NY&USA&2.61&1.8&1.9&2.0&93.3&172.3&183.6&1300.5&2547.7&1560.4 \\
Charleston, OR&USA&1.18&37.7&53.8&70.5&2505.9&3108.4&4800.4&13168.6&20908.5&19984.6 \\
South Beach, OR&USA&1.37&22.2&31.1&38.2&2037.2&2469.1&3720.7&12483.4&19973.4&18772.7 \\
Astoria, OR&USA&1.25&9.9&15.3&20.0&1179.6&1421.2&2048.2&6321.3&10045.6&9001.4 \\
Newport, RI&USA&1.46&8.3&11.7&12.4&1572.1&3427.3&3138.9&8996.1&22403.0&16090.1 \\
Charleston, SC&USA&1.61&3.0&3.2&3.6&835.1&1508.4&1703.2&7925.2&15785.5&13080.1 \\
Galveston (Pier 21), TX&USA&2.96&1.9&1.9&1.9&33.4&155.6&138.6&1648.4&3005.7&1985.6 \\
Rockport, TX&USA&1.97&2.1&2.2&2.3&510.5&923.3&913.8&7270.5&15475.8&11903.8 \\
Port Isabel, TX&USA&2.76&1.3&1.3&1.3&7.5&126.3&115.7&1251.6&2014.6&1396.5 \\
Galveston (P. Pier), TX&USA&5.26&1.3&1.2&1.3&1.6&1.7&1.7&152.5&211.6&145.1 \\
Chesapeake BBT, VA&USA&2.02&3.4&3.7&4.2&461.4&909.9&895.4&4740.5&10662.9&7203.3 \\
Neah Bay, WA&USA&1.31&4.2&7.1&10.8&686.8&818.8&1061.3&3497.6&5115.0&4501.7 \\
Willapa Bay, WA&USA&2.05&2.2&2.6&2.8&251.3&269.6&382.2&2185.3&3014.1&2590.3 \\
Lewes, DE&USA&2.18&2.5&2.6&2.9&269.9&551.6&557.6&2789.4&6303.5&4022.3 \\
Apra Harbor, Guam&USA Trust&0.50&2417.4&3602.8&3793.9&51797.9&62775.5&70531.0&69295.5&75780.4&78992.1 \\
Wake&USA Trust&1.82&1.5&1.6&1.7&618.0&927.2&813.2&5042.8&8596.8&8543.8 \\
Johnston&USA Trust&0.89&45.7&171.4&76.8&11344.6&20089.5&22332.8&35174.6&47531.8&49323.9 \\
Midway&USA Trust&1.12&7.5&16.1&14.8&3714.3&6376.9&9526.7&22321.2&34381.6&34488.4 \\
Pago Pago&USA Trust&0.48&7167.3&11780.5&12692.2&75498.9&81710.1&82693.4&82939.5&86495.5&86097.5 \\
Charlotte Amalie, VI&USA Trust&1.24&1.8&1.7&1.8&1590.1&2712.5&2494.8&13142.3&23022.3&22474.6 \\
San Juan, PR&USA Trust&0.93&4.4&8.7&5.7&5947.9&11394.3&11413.4&31020.1&44208.2&46298.4 \\
Magueyes Island, PR&USA Trust&1.09&1.9&1.9&2.0&2508.2&4807.1&4466.8&18458.3&30861.5&31125.9 \\
\hline
\end{longtable}}
\label{Stab:af_table_rp500}
\end{landscape}

}

\end{document}